\documentclass[aps,prd, onecolumn,showpacs,showkeys,groupedaddress,11pt]{revtex4}
\usepackage{amsmath, amsthm, amssymb, amsfonts}
\usepackage{graphicx} 
\usepackage{color}
\usepackage{wrapfig}
\usepackage{float}
\usepackage[normalem]{ulem}
\usepackage{soul}

\begin{document}

\title{Instanton solutions on the polymer harmonic oscillator}
\author{Joan A. Austrich Olivares}
\email{jascao90@hotmail.com}
\author{Angel Garcia-Chung}
\email{angel.garcia@correo.nucleares.unam.mx} 
\author{J. David Vergara}
\email{vergara@nucleares.unam.mx} 
\affiliation{Instituto de Ciencias Nucleares, Universidad Nacional Aut\'onoma de M\'exico, A. Postal 70-543, M\'exico D.F., 04510, MEXICO}

\date{\today}

\begin{abstract}
It is computed, using instanton methods, the first allowed energy band for the polymer harmonic oscillator. The result is consistent with the band structure of the standard quantum pendulum but with pure point spectrum. An effective infinite degeneracy emerges in the formal limit $\mu/l_0 \to 0$ where $l_0$ is the characteristic length of the vacuum eigenfunction of a quantum harmonic oscillator. As an additional result, it is shown along the article the role played by the lattice reference point $\lambda$ in the full quantization of the polymer harmonic oscillator.
\end{abstract}

\maketitle

%

\section{Introduction.}

The nonstandard polymer quantization \cite{ShadowsStates, PQMCLimit, HPHSPQM} is a quantization procedure for mechanical models which implements some of the techniques of the loop quantization program \cite{LQProgram1, LQProgram2, LQProgram3}. Its main feature is a singular representation of the Weyl algebra \cite{Bratelli} in a non-separable Hilbert space and as a result, the canonical commutation relations \cite{vonNeumann} cannot be recovered by using the Stone-von Neumann theorems \cite{Strocchi1, Strocchi2, Halvorson}. 

Polymer quantization can be applied to a broad range of mechanical systems related with atomic physics \cite{Anakin, ViqarLouko, Louko1, Louko2, Barbero} as well as  in cosmology \cite{Abhay, Martin} or in quantum field theory \cite{Viqar} and statistical mechanics \cite{Memo1, Memo2, Memo3} just to mention some examples. Each of them serves as a toy model which provides specific hints about the procedure required to compare the non-regular and regular quantizations in the much more complicated context of quantum space-time. Nevertheless, on its own it represents an interesting model to study from a mathematical physics perspective \cite{HugoyYo}.

The non-regularity of the Weyl algebra representation avoids the definition of the momentum operator and makes impossible the usual dynamical characterization of the system. This obstacle is circumvented by proposing an operator that mimics the role of the absent operator \cite{ShadowsStates, PQMCLimit}. This artificial operator is formed as a combination of holonomies $\widehat{V}_\mu$ and depends on a length parameter $\mu$. For the case of the polymer harmonic oscillator, the standard kinematical term $\widehat{p}^2$ is replaced by the operator 
\begin{equation}
 {\widehat{P^2}^{(\mu)}_{eff}} := \frac{\hbar^2}{2m \mu^2} \left[ 2 - \widehat{V}_\mu - \widehat{V}_{-\mu} \right], \label{KTerm}
 \end{equation}
\noindent where $\widehat{V}_\mu$ is the `holonomy operator' with a fixed value for the parameter $\mu$. 

 In the loop quantum gravity (LQG) scenario, the analogous of the parameter $\mu$ possesses a fixed value given by the Planck length. For mechanical models, the nature of the parameter is ambiguous due to its non-fundamental characterization. However, its value is fixed although undetermined and can be seen as providing a measure of discretization of space. Since no discreteness of space has been detected, the deviations of the polymer physical quantities from those observed within the regular quantization cannot be experimentally tested. On the other hand, the physical quantities involved in polymer systems acquire corrections which are proportional to (powers of) $\mu$. As a consequence, in mechanical systems $\mu$ must be much smaller than the natural length parameters associated with the system. The estimated value of the parameter $\mu$ for the free particle and the harmonic oscillator is about $\mu \sim 10^{-19}m$ \cite{ShadowsStates, PQMCLimit, Anakin, Viqar}. The quantum corrections of these systems become significant in regimes where quantum mechanics would be inapplicable. 
 
  Hamiltonian commutation relations between the operator $\widehat{x}$ and an arbitrary holonomy $\widehat{V}_{\alpha}$ are given by $[\widehat{x}, \widehat{V}_\alpha] =- \alpha \widehat{V}_\alpha$. They are irreducibly represented on the Hilbert space ${\cal H}_{poly}$ which is given by $L^2(\overline{\mathbb{R}}, d\mu_0)$. Here,  $\overline{\mathbb{R}}$ is the Bohr compactification of the real line and $d\mu_0$ is a regular Haar measure on it \cite{Bohr}. This Hilbert space was built by  mimicking the loop quantization program as was mentioned before \cite{ShadowsStates, PQMCLimit, Velhinho}. In this case, the representation of the operators $\widehat{x}$ and $\widehat{V}_{\alpha}$ is
\begin{equation}
\widehat{x} N_\beta (p) = i\hbar \frac{d}{dp}N_\beta(p), \qquad \widehat{V}_\alpha N_\beta(p) = N_{\alpha + \beta} (p),
\end{equation}
\noindent where $N_\beta(p):= e^{i \frac{\beta p}{\hbar}}$ is an element of the uncountable basis on ${\cal H}_{poly}$, i.e., $\beta \in \mathbb{R}$.

A particular effect of the polymer construction coming as a result of the introduction of the aforementioned parameter is the modification of the dynamical equation of the quantum system. The Hamiltonian of the system when using the kinetic term (\ref{KTerm}) takes the form
\begin{equation}
\widehat{H}^{(\mu)}_{poly} = \frac{\hbar^2}{2m \mu^2} \left[ 2 - \widehat{V}_\mu - \widehat{V}_{-\mu} \right] + \frac{1}{2} m \omega^2 \widehat{x}^2. \label{PHamiltonian}
\end{equation}
Hamiltonian (\ref{PHamiltonian}) splits the full polymer Hilbert space ${\cal H}_{poly}$ as an (uncountable) direct sum of separable Hilbert spaces ${\cal H}^{(\lambda)}_{poly}$ of the  form ${\cal H}_{poly} = \oplus_{\lambda \in [0,\mu)} {\cal H}^{(\lambda)}_{poly}$ or in the notation used in \cite{ReedSimon}, ${\cal H}_{poly} = \int^{\oplus}_{\lambda \in [0,\mu)} {\cal H}^{(\lambda)}_{poly} \, d\lambda^c$ where $d\lambda^c$ is the countable measure on the interval $[0, \mu)$. Consequently, the non-separability of the Hilbert space ${\cal H}_{poly}$ renders discrete the nature of the parameter $\lambda$. Each ${\cal H}^{(\lambda)}_{poly}$ is given by ${\cal H}^{(\lambda)}_{poly} = L^2([-\pi \frac{\hbar}{\mu}, + \pi \frac{\hbar}{\mu}),dp)$ and their observables algebra is now generated by $\widehat{x}$ and $\widehat{V}_{n \mu}$ where $n \in \mathbb{Z}$. As a result, these Hilbert spaces ${\cal H}^{(\lambda)}_{poly}$, each of them  labeled by $\lambda$, are superselected. Notice that the spatial scale given by $\mu$ naturally induces a scale with momentum dimensions given by 
 \begin{equation}
 p_c = \hbar/\mu \approx 10^{-15} m \, kg \, s^{-1}. \label{MChac}
 \end{equation}
 The Schr\"odinger equation for a given state $\Psi^{(\lambda)}(p) \in {\cal H}^{(\lambda)}_{poly}$ takes the form
\begin{equation}
-\frac{m \hbar^2 \omega^2 }{2}\frac{d^2}{dp^2} \Psi^{(\lambda)}(p) + \frac{\hbar^2}{m \mu^2}\left[ 1 -  \cos\left( \frac{p}{p_c} \right)  \right] \Psi^{(\lambda)}(p) = E^{(\lambda)} \Psi^{(\lambda)}(p), \label{SchrP}
\end{equation}
\noindent with the additional condition
 \begin{equation}
 \Psi^{(\lambda)}(p + 2\pi p_c) = e^{2 \pi i \frac{p_c \lambda}{\hbar}}  \Psi^{(\lambda)}(p), \label{BCondition}
 \end{equation}
\noindent  in order to properly generate states in the entire polymer Hilbert space using states of each superselected sector. Observe that when $\lambda = 0$ in (\ref{BCondition}), then the wave functions $\Psi^{(\lambda =0)}(p)$ are $2 \pi p_c$ periodic functions. On the other hand, the parameter $\lambda$ in (\ref{BCondition}) can be seen as a fixed dual Bloch wave vector. 

\begin{figure}[!ht]  
    \includegraphics[width=8cm]{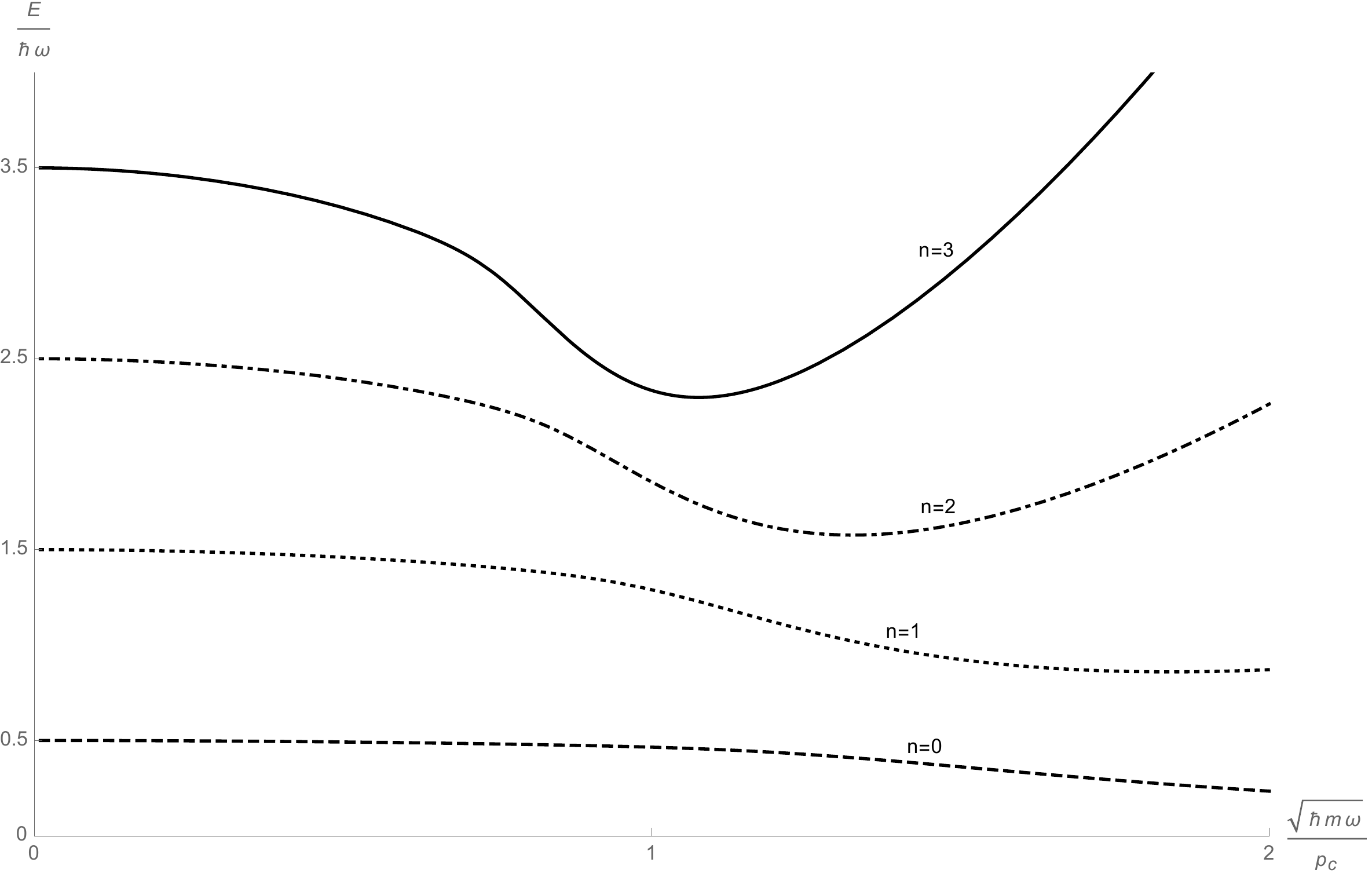} 
  \caption{Eigenvalues $n= \{ 0, 1, 2, 3 \}$ in terms of the quotient $\sqrt{\hbar m \omega}/p_c $.} \label{EigenValues}
\end{figure}

  The equation (\ref{SchrP}) can be written into the more familiar form of the Mathieu equation \cite{Abramovitz} by using a dimensionless variable $u := \frac{1}{2}(\frac{p}{p_c} + \pi)$ within the interval $u \in [0, \pi)$ as
\begin{equation}
\frac{d^2}{du^2}\Psi^{(\lambda)}(u) + \left( b - 2 q \cos(2u) \right)\Psi^{(\lambda)}(u) = 0, \label{MEquation}
\end{equation}
\noindent where the parameters $b$ and $q$ in (\ref{MEquation}) are given by
\begin{equation}
b:= \frac{8 p^2_c}{m \omega^2 \hbar^2} \left( E - \frac{p^2_c}{m}\right), \qquad q := \left( \frac{2 p^2_c}{m \omega \hbar}\right)^2. \label{MEqPara}
\end{equation}
The spectrum of the Eq. (\ref{MEquation}) is given by a series derived with a recursive method and the appropriate boundary conditions \cite{Abramovitz}. Consequently, the expression for any of the eigenvalues of the Mathieu equation are very difficult to handle at analytical level, and hence, only its approximate expressions can be used. If we consider periodic solutions (i.e., $\lambda =0$) then the spectrum of the Mathieu equation \cite{Abramovitz} given by (\ref{MEquation}) tends to the spectrum of the quantum harmonic oscillator in the formal limit $\mu \rightarrow 0$ (Figure \ref{EigenValues}). In this limit, the known results of the standard theory of the quantum harmonic oscillator are recovered as we already mentioned. For cases in which $\lambda \neq 0$, the equation can be solved by using Floquet theory for non-periodic Mathieu functions \cite{Abramovitz}. 

 As can be seen, an interesting result of the replacement of the kinetic term given in (\ref{KTerm}) is that in the momentum representation, this term turns out to be a periodic potential, in sharp contrast to the quantum harmonic oscillator in which this property is absent. Nevertheless, it also renders periodic the semiclassical description, via path integral formulation \cite{Campiglia, ParraDavid, Hugo} which is, in this sense, a desired result. This feature results from the fact that (\ref{KTerm}) is a particular case of an almost periodic Schr\"odinger operator \cite{Cycon, Araki} i.e., operators formed by a finite sum of holonomies.
 
 Recently, Barbero et al. \cite{Barbero} showed that the energy spectrum of (\ref{PHamiltonian}) consists of an uncountable number of eigenvalues grouped in bands, very much like those appearing in the study of periodic potentials in standard quantum mechanics. The main difference is that the polymer energy spectrum is entirely discrete although grouped in bands whereas in the standard periodic potentials is continuum and grouped in bands. This peculiarity of the spectrum creates some difficulties in the development of the formalism both to define unitary evolution and to build suitable statistical ensembles. It also emerges both, at the full loop theory and, in its symmetry reduced quantum mechanical versions \cite{Abhay, Martin}. However, some proposals for constructing a separable Hilbert space are recently given for the case of quantum mechanical models \cite{Tomasz1, Tomasz2, Barbero2}. They essentially propose replacing the countable measure $d\lambda^c$ by a Lebesgue measure $d\lambda$ on the same interval. 

 A key point in Barbero's results is the Burnat-Shubin-Herczy\'nski theorem \cite{TheoremSH}. This theorem establishes that the spectrum of an almost periodic operator is the same no matter if the representation is regular or singular. Representations only modify the nature of the spectrum of almost periodic Schr\"odinger operators in that if it is discrete or continuous. In the case of the Hamiltonian (\ref{PHamiltonian}), its spectrum is identical to that of the quantum pendulum in the standard representation \cite{Barbero, BSimon}.
 
 Periodic potential systems are very familiar in standard quantum mechanics, particularly in solid state physics \cite{ReedSimon, Kittel}. Among other features, these potentials have an infinite set of oscillators-like eigenstates only when barrier penetration is ignored. Otherwise, tunneling effect between each of the `classical vacuums' gives rise to the band structure of the spectrum. In standard quantum mechanics, the tunneling effect can be realized by instanton and anti-instanton pseudo-particles in the long imaginary time regime \cite{Coleman, Rajamaran, ABCInstanton, Kazama} although it may be obtained within the deep quantum regime \cite{Paradis}. Instanton methods are a non-perturbative technique developed for handling phenomena in quantum field theory, and in particular in quantum chromodynamics, analogous to barrier penetration in quantum particle mechanics. 
  
 In the polymer quantization, the tunnel effect can also yields the band structure of the spectrum indicated by Barbero et al. In this scenario, the long time behavior of the penetration amplitude of the polymer harmonic oscillator by using instanton methods can be considered. In this case, each well at $p = p_c (2n-1) \pi$ entails a classical vacuum from which the tunneling effect can occur in analogy to the result in the standard quantum pendulum \cite{BSimon}. Having said this, our goal here is to determine the penetration amplitude of the tunneling effect on the polymer harmonic oscillator between the minima in the entire polymer Hilbert space and to obtain the band structure of the minimum of the spectrum. For this purpose, we are going to apply instanton methods to the polymer harmonic oscillator.
 
 It is worth mentioning that the mathematical structure of the polymer harmonic oscillator (and other polymer quantum systems) is very much similar to that of the lattice quantization of such systems \cite{Chalbaud}. Notice for example that the super-selected Hilbert space ${\cal H}^{(\lambda)}_{poly}$ is the same Hilbert space used in the lattice quantization of the harmonic oscillator \cite{Chalbaud}. However, although they coincides in this Hilbert space, in both schemes there are conceptual aspects which makes them considerably different. For example, the polymer quantum mechanics (PQM), as mimicking the LQG procedure, can be seen as a result of the GNS-construction of the Weyl algebra with a non-regular positive linear functional \cite{PQMCLimit, Velhinho}.  The discreteness of the space in PQM is a result of such positive linear functional (see \cite{ShadowsStates} for details) and there is not a preferred lattice or so involved at this point. The lattice-type similarity emerges when the fictitious operator $ {\widehat{P^2}^{(\mu)}_{eff}}$ given in (\ref{KTerm}) is invoked. As a result, the fundamental discreteness of the space is unveiled by the parameter $\mu$. For this reason, the limit $\mu \rightarrow 0$ can be seen as a formal mathematical trick in order to compare the physical quantities within the polymer description with their similar in the standard quantization.
 
 On the other hand, lattice quantum mechanics is a generalization of the standard quantization techniques on a discrete space. Its powerful cousin, the lattice quantum field theory, provides remarkable results in removing the awkward divergences of the quantum field theory within a continuous space. Such results has proven to be particularly useful in quantum chromodynamics \cite{Wilson, Jan}. However, in these cases, the lattice structure together with the discreteness of the space are fixed by hand, that is to say, the lattice is not a fundamental description of the space. In order to recover the physical description of the system, the limit when the lattice parameter tends to zero is the ultimate step.
 
 Summarizing, both schemes coincide mathematically once the super-selected sector in polymer quantum mechanics is fixed. The differences between both quantum models are twofold: the non-separability of the Hilbert space which is a distinguishing feature of the polymer quantum mechanics, whereas in lattice quantum mechanics the Hilbert space is separable. Secondly, the limit of the lattice parameter $\mu \rightarrow 0$, which is a mathematical trick in polymer quantum mechanics, is the final procedure in lattice quantization. 
  
 In this spirit, the present work is directed to establish a link between the known results of lattice quantum mechanics and instanton methods applied to the quantum pendulum and the polymer quantization of the harmonic oscillator. The discrete nature of the parameter $\lambda$ in these results offers new insights for instance in the statistical mechanics of the polymer harmonic oscillator and in the polymer quantization of the scalar field as we will discuss in the last section. 

The paper is organized as follows. In Section \ref{PHO} we express the transition amplitude of the polymer harmonic oscillator in terms of the amplitudes of non-restricted regular Hilbert spaces. In Section \ref{InstantonSol} we obtain the instanton solutions and discuss its main properties. The analysis of the quantum fluctuation of the instanton solution together with the penetration barrier amplitude calculation is reported in Section \ref{QFInst}.  Finally in Section \ref{Discussion} we discuss our results.


\section{Propagator of the Polymer Harmonic Oscillator(PHO)} \label{PHO}


 Let us consider an arbitrary state $\Psi(p) \in {\cal H}_{poly}$. This state can be decomposed as $\Psi(p) = \oplus_{\lambda} \Psi^{(\lambda)}(p)$, where $\Psi^{(\lambda)}(p) \in {\cal H}^{(\lambda)}_{poly}$. As we already mentioned, the Hilbert spaces ${\cal H}^{(\lambda)}_{poly}$ are superselected and therefore, the Hamiltonian $\widehat{H}^{(\mu)}_{poly}$ given in (\ref{PHamiltonian}) acts on this state as
\begin{equation}
\widehat{H}^{(\mu)}_{poly} \left( \oplus_{\lambda} \Psi^{(\lambda)}(p) \right) =  \left( \oplus_{\lambda} \widehat{H}^{(\mu)}_{poly} \Psi^{(\lambda)}(p) \right) ,
\end{equation}
\noindent and notice that the Hamiltonian operator moves inside the summation on the right hand side. As a consequence, the evolution operator of a given state $\Psi^{(\lambda)}(p)$ takes the form
\begin{equation}
\langle p_f , t_f | p_i, t_i \rangle^{(\lambda)}  =  \langle p_f | e^{- \frac{i}{\hbar} (t_f - t_i) \widehat{H}^{(\mu)}_{poly} } | p_i \rangle^{(\lambda)} , \label{Amplitude}
\end{equation}
\noindent where the Hamiltonian operator is now understood as the one acting on the Hilbert space ${\cal H}^{(\lambda)}_{poly}$. From now on let us fix the value of $\lambda$ and let us make all the calculations on this specific Hilbert space ${\cal H}^{(\lambda)}_{poly}$. The symbol $\lambda$ will be omitted for simplicity unless it is required to avoid confusion. 

The momentum variable $p$ in the quantum configuration space of ${\cal H}^{(\lambda)}_{poly}$ is confined to the interval $p \in [-\pi p_c, + \pi p_c)$ and this is a topological constraint \cite{Kleinert}. In this section, we obtain an expression (\ref{Amplitude}) in terms of the amplitude of a system without the topological constraint, i.e., a system in which the momentum variable $p \in \mathbb{R}$. To do so, let us follow the procedure given in \cite{Kleinert} for these systems. 

 First, let us decompose the time interval $t_f - t_i$ given in (\ref{Amplitude}) into $N+1$ pieces $\epsilon := (t_f - t_i)/(N+1) = t_j - t_{j-1}$, where $t_0 := t_i$ and $t_{N+1} := t_f$. The exponential in (\ref{Amplitude}) can now be written as $N+1$ products of infinitesimal time interval exponentials
\begin{equation}
\langle p_f , t_f | p_i, t_i \rangle  = \langle p_f | e^{ -\frac{i(t_f - t_N)}{\hbar}  \widehat{H}^{(\mu)}_{poly} } e^{ -\frac{i(t_N - t_{N-1})}{\hbar}  \widehat{H}^{(\mu)}_{poly} } \cdots e^{ -\frac{i(t_1 - t_i)}{ \hbar}  \widehat{H}^{(\mu)}_{poly} } | p_i \rangle =   \langle p_f | \prod^{N+1}_{j=1} e^{- \frac{i \epsilon}{\hbar}  \widehat{H}^{(\mu)}_{poly} } | p_i \rangle .
\end{equation}
 
 In the polymer construction, within the $p-$polarization, the completeness relation of the momentum basis \cite{PQMCLimit, HPHSPQM, Anakin} is of the form
 \begin{equation}\label{completep}
  \frac{1}{2 \pi p_c} \int^{+ \pi p_c }_{- \pi p_c } dp \, |p\rangle \langle p | = \widehat{1} .
  \end{equation} 
Notice that the integration must end at an infinitesimal piece below $+ \pi p_c$ to avoid the double-counting of contributions of the identical points $p=- \pi p_c$ and $p=+ \pi p_c$.  We now introduce (\ref{completep}) on each product and obtain
\begin{equation}
\langle p_f , t_f | p_i, t_i \rangle  =  \left[ \prod^{N}_{n=1} \int^{+\pi p_c}_{- \pi p_c} \frac{dp_n}{2 \pi p_c} \right] \prod^{N+1}_{j=1} \langle p_j , t_j | p_{j-1}, t_{j-i} \rangle , \label{TDecomp}
\end{equation} 
\noindent where $p_0:= p_i$ and $p_{N+1}:= p_f$. This expression allow us to derive separately the infinitesimal amplitudes $\langle p_j , t_j | p_{j-1}, t_{j-i} \rangle$ which can be written as
\begin{equation}
 \langle p_j , t_j | p_{j-1}, t_{j-i} \rangle =  \langle p_j  | e^{- \frac{i \epsilon}{ \hbar} \widehat{H}^{(\mu)}_{poly} } |p_{j-1} \rangle, \label{FInfiAm}
\end{equation}
\noindent and in order to calculate (\ref{FInfiAm}) we consider apart the zeroth Hamiltonian infinitesimal amplitude 
\begin{equation}
 \langle p_j , t_j | p_{j-1}, t_{j-i} \rangle^{(0)} =  \langle p_j  |p_{j-1} \rangle. \label{SAmplitude}
\end{equation}

In an unconstrained system, the right hand side of the previous expression is the Dirac delta orthonormality condition of the basis elements $| p \rangle $, that is to say, 
$  \langle p  |p' \rangle = 2 \pi p_c \, \delta(p - p' )$. In the present case, we cannot use this relation because is not well defined. The reason is that it gives rise to a boundary condition different to that in (\ref{BCondition}). For our case, the adequate orthonormality condition in ${\cal H}^{(\lambda)}_{poly}$ reads as
\begin{equation}
\langle p  | p' \rangle = 2 \pi p_c \sum_{n \in \mathbb{Z}} e^{2\pi  i n \lambda/\mu } \delta(p - p' - 2 \pi n p_c). \label{DriacDIn}
 \end{equation}
 
To verify that we recover the relation (\ref{BCondition}) from the relation (\ref{DriacDIn}), let us add $2\pi p_c$ to the $p$ variable on both sides of (\ref{DriacDIn}). After some algebraic manipulations we obtain
$$ \langle p +2 \pi p_c | p' \rangle =  e^{2\pi i \frac{p_c \lambda}{\hbar}} \langle p | p' \rangle ,$$
\noindent and now, removing $| p' \rangle $ from both sides and multiplying by $|\Psi \rangle$ we recover (\ref{BCondition}).

 Using (\ref{DriacDIn}) the zeroth Hamiltonian amplitude in (\ref{SAmplitude}) can be written as
\begin{eqnarray}
 \langle p_j , t_j | p_{j-1}, t_{j-i} \rangle^{(0)} &=&  2 \pi p_c \sum_{n \in \mathbb{Z}} e^{2\pi  i n \lambda/\mu } \delta(p_j - p_{j-1} - 2 \pi n p_c) \nonumber \\
 &=& \frac{1}{2} \sum_{n \in \mathbb{Z}} e^{ 2 \pi i n \frac{\lambda}{\mu}} \int^{+\infty}_{-\infty} d\varphi_j \, e^{ i \frac{\varphi_j}{2 p_c} ( p_j - p_{j-1} - 2 \, \pi n p_c) }, \label{ZHamil} 
\end{eqnarray}
\noindent where the Dirac deltas inside the summation in the first line are written in terms of its Fourier transforms in the second line. We now use the spectral decomposition formula \cite{Kleinert} to obtain the infinitesimal amplitude for the Hamiltonian $ \widehat{H}^{(\mu)}_{poly}$ 
\begin{equation}
 \langle p_j , t_j | p_{j-1}, t_{j-i} \rangle =  \langle p_j |e^{ -\frac{i \epsilon}{\hbar} \widehat{H}^{(\mu)}_{poly} } | p_{j-1} \rangle = \frac{1}{2} \sum_{n \in \mathbb{Z}} e^{ 2 \pi i n \frac{\lambda}{\mu}} \int^{+\infty}_{-\infty} d\varphi_j \, e^{ i \frac{\varphi_j}{2 p_c} ( p_j - p_{j-1} - 2 \, \pi n p_c)  - \frac{i \epsilon}{\hbar} H^{(\mu)}_{poly}(p_j, \frac{ \mu \varphi_j}{2}) }. \label{InfinAmplitude}
\end{equation}
\noindent The Hamiltonian $H^{(\mu)}_{poly}(p_j,x_j)$ is given by
\begin{equation}
H^{(\mu)}_{poly}(p_j, x_j = \mu \varphi_j/2) := \frac{2\hbar^2}{m \mu^2} \sin^2\left( \frac{p_j}{2 p_c} \right) + \frac{k \mu^2}{8} \varphi^2_j,
\end{equation}
\noindent where $k = m \omega^2$. After inserting the amplitude (\ref{InfinAmplitude}) in the expression (\ref{TDecomp}) we arrived at the following expression
\begin{eqnarray}
\langle p_f , t_f | p_i, t_i \rangle  =  \prod^{N}_{n=1} \int^{+\pi p_c}_{- \pi p_c} \frac{dp_n}{2 \pi p_c} \left[  \prod^{N+1}_{j=1}  \sum_{n_j \in \mathbb{Z}} \int^{+\infty}_{-\infty} \frac{e^{ 2 \pi i n_j \frac{\lambda}{\mu}}\,d\varphi_j}{2} \, e^{ i \frac{\varphi_j}{2 p_c} ( p_j - p_{j-1} - 2 \, \pi n_j p_c)  - \frac{i \epsilon}{\hbar} H^{(\mu)}_{poly}(p_j, \frac{ \mu \varphi_j}{2}) } \right].
\end{eqnarray}

On each of the $N$ integrals $dp_n$, the momenta variables $p_n$ can be redefined by allowing them to go to a broader range $\tilde{p}_j \in (-\infty, +\infty)$ and absorbing one particular summation (see the Appendix \ref{PATHSUM} for details). As a result, $N$ summations will be absorbed by $N$ integrations and therefore, there is only one summation left in this procedure.  The total amplitude takes the following form
\begin{eqnarray}
\langle p_f , t_f | p_i, t_i \rangle  &=&  \left(  \frac{2\pi \hbar}{  i \epsilon k \mu^2 }  \right)^{(N+1)/2} \sum_{l \in \mathbb{Z}} e^{2 \pi i l \frac{\lambda}{\mu}} \, \left[ \prod^{N}_{n=1} \int^{+\infty}_{-\infty} \frac{d \tilde{p}_n}{2 \pi p_c } \right] \exp\left\{ \frac{i \epsilon }{ \hbar} \sum^{N+1}_{j=1} \left[  - \frac{2 p^2_c }{m} \sin^2(\frac{\tilde{p}_j}{2 p_c}) +  \right.  \right. \nonumber \\
&& \left.  \left.   + \frac{1}{2k} \left( \frac{ \tilde{p}_j - \tilde{p}_{j-1} + 2 \pi p_c l \delta_{j, N+1} }{\epsilon} \right)^2                     \right]  \right\} . \label{AmplitudePoly}
\end{eqnarray} 

The potential $V(\tilde{p}_j) :=\frac{2 p^2_c}{m} \sin^2(\frac{\tilde{p}_j}{2p_c})$ is indeed periodic and invariant under this redefinition of the variables $\tilde{p}_j$. Notice that for each value $l \in \mathbb{Z}$ we obtain an amplitude
\begin{eqnarray}
\langle \langle p_f + 2 \pi p_c l, t_f | p_i, t_i \rangle \rangle  &:=&  \left(  \frac{2\pi \hbar}{  i \epsilon k \mu^2}  \right)^{(N+1)/2} \left[ \prod^{N}_{n=1} \int^{+\infty}_{-\infty} \frac{d \tilde{p}_n}{2 \pi p_c } \right] e^{ \frac{i \epsilon }{ \hbar} \sum^{N+1}_{j=1}\left[ \frac{1}{2k} \left( \frac{ \tilde{p}_j - \tilde{p}_{j-1}  }{\epsilon} \right)^2 - V(p_j) \right] }, \nonumber \\
& = & {\cal N} \int^{p_f + 2 \pi p_c l}_{p_i}  \frac{{\cal D} p}{2 \pi p_c}\,  e^{ \frac{i }{ \hbar} \int^{t_f}_{t_i} dt \left[ \frac{1}{2k} \dot{p}^2 - V(p) \right] }, \label{PLPendulo}
\end{eqnarray} 
\noindent in agreement to the general derivation given in \cite{Kleinert}. The symbol $\langle \langle \cdot | \cdot \rangle \rangle $ in this amplitude indicates that the amplitude corresponds to that of a unconstrained topological system, i.e., the domain of the momentum variable comprised the entire real line $ p \in (-\infty ,+ \infty)$. Finally, inserting (\ref{PLPendulo}) in (\ref{AmplitudePoly}) we obtain the amplitude of the polymer harmonic oscillator in the Hilbert space ${\cal H}^{(\lambda)}_{poly}$
\begin{equation}
\langle p_f , t_f | p_i, t_i \rangle = \sum_{n \in \mathbb{Z}} e^{2 \pi i n \frac{\lambda}{\mu}}  \langle \langle p_f + 2 \pi p_c n, t_f | p_i, t_i \rangle \rangle. \label{PolyAmplitudeSum}
\end{equation}

From (\ref{PLPendulo}) we can extract out the effective action in momentum variables
\begin{equation}
S[p] := \int^{t_f}_{t_i} dt \left[ \frac{1}{2k} \dot{p}^2 - V(p) \right] , \label{PAction}
\end{equation}
\noindent with momentum $p$ as the dynamical variable instead of the usual $x$ and the potential $V(p)$ is given by 
\begin{eqnarray}
 V(p) :=  \frac{2 p^2_c}{ m } \sin^2\left( \frac{p}{2 p_c} \right). \label{PolyPot}
\end{eqnarray}
Clearly, this potential is periodic with period $2 \pi p_c$ and its action (\ref{PAction}) corresponds with the action of a simple pendulum in the momentum space.

The amplitudes appearing in the right hand side of (\ref{PolyAmplitudeSum}) can be regularized if we consider the amplitude of the standard harmonic oscillator
\begin{eqnarray}
 \langle p_f, t_f | p_i, t_i \rangle^{(H)}  &=&  \left(  \frac{\hbar}{ 2\pi  i \epsilon k x^2_0 }  \right)^{(N+1)/2} \left[ \prod^{N}_{n=1} \int^{+\infty}_{-\infty} \frac{d p_n}{2 \pi p_0 } \right] e^{ \frac{i \epsilon }{ \hbar} \sum^{N+1}_{j=1}\left[ \frac{1}{2k} \left( \frac{ \tilde{p}_j - \tilde{p}_{j-1}  }{\epsilon} \right)^2 - V_0(p_j) \right] }, \nonumber \\
& = & {\cal N}_0 \int^{p_f}_{p_i}  \frac{{\cal D} p}{ 2 \pi p_0} \,  e^{ \frac{i }{ \hbar} \int^{t_f}_{t_i} dt \left[ \frac{1}{2k} \dot{p}^2 - V_0(p) \right] }, \label{amplitudeHO}
\end{eqnarray}
\noindent where the parameters $x_0$ and $p_0$ were introduced in order to make the measure of the integral dimensionless as well as the factor ${\cal N}_0$. In our next analysis we will fix the values of $x_0$ and $p_0$ in terms of $\mu$ and $p_c$. In the amplitude (\ref{amplitudeHO}), $V_0(p)$ is the potential of the harmonic oscillator in momentum variables $V_0(p) = p^2/2m$. 

Let us multiply and divide by the standard Harmonic oscillator amplitude, given in (\ref{amplitudeHO}), each summation term on the right hand side of (\ref{PolyAmplitudeSum})
\begin{eqnarray}
\langle \langle p_f, t_f | p_i, t_i \rangle \rangle =  \langle p_f, t_f | p_i, t_i \rangle^{(H)}  \frac{ {\cal N} \int^{p_f}_{p_i}  \frac{{\cal D} p\,  e^{ \frac{i }{ \hbar} \int^{t_f}_{t_i} dt \left[ \frac{1}{2k} \dot{p}^2 - V(p) \right] } }{2 \pi p_c}}{{\cal N}_0 \int^{p_f }_{p_i}  \frac{{\cal D} p\,  e^{ \frac{i }{ \hbar} \int^{t_f}_{t_i} dt \left[ \frac{1}{2k} \dot{p}^2 - V_0(p) \right] }}{ 2 \pi p_0}  } . 
\end{eqnarray}
\noindent  In order to remove the divergence associated with the quotient ${\cal N}/{\cal N}_0 $, we fix the values of the parameters $x_0 = 2 \pi \mu$ and $p_0 = \frac{p_c}{2 \pi} $. This yields ${\cal N}/{\cal N}_0 =1$ and gives the amplitude the form
\begin{eqnarray}
\langle \langle p_f, t_f | p_i, t_i \rangle \rangle &=& \langle p_f, t_f | p_i, t_i \rangle^{(H)}  \frac{ \int^{p_f}_{p_i}  {\cal D} p \,  e^{ \frac{i }{ \hbar} \int^{t_f}_{t_i} dt \left[ \frac{1}{2k} \dot{p}^2 - V(p) \right] } }{ \int^{p_f}_{p_i}  {\cal D} p \,  e^{ \frac{i }{ \hbar} \int^{t_f}_{t_i} dt \left[ \frac{1}{2k} \dot{p}^2 - V_0(p) \right] } } . \label{RegAmplitude}
\end{eqnarray}

 Now that we have derived the penetration amplitude of the polymer harmonic oscillator in terms of the amplitude of the harmonic oscillator in momentum variables $\langle p_f, t_f | p_i, t_i \rangle^{(H)}$, the next step is to approximate the quotient of integrals in (\ref{RegAmplitude}). To do so, each action on (\ref{RegAmplitude}) is expanded around their corresponding instanton solution $P$. The expansion will be carried out up to second order in the quantum fluctuations $\delta p$ \cite{Coleman, Rajamaran, ABCInstanton, Kazama}. The purpose of the next section is the calculation of the Instanton solution $P^+$ of the classical action (\ref{PAction}).


\section{Instanton solution of the polymer harmonic oscillator} \label{InstantonSol}


In this section, we study the instanton solution associated with the action (\ref{PAction}). Instantons are solutions of the classical equation of motion in imaginary time, i.e., on equations written after a Wick rotation of the time parameter $t =-i \tau$. The Wick rotation of the action (\ref{PAction}) gives the following Euclidean action 
\begin{eqnarray}
 {\cal S}_E &=& \int^{\tau_f}_{\tau_i} d\tau  \left[ \frac{{p'}^2}{2k}  + V(p) \right] = \int^{\tau_f}_{\tau_i} d\tau  \left[ \frac{{p'}^2}{2k}  + \frac{2 p^2_c}{ m} \sin^2\left( \frac{ p}{2 p_c} \right) \right] , \label{MLagrangian}
\end{eqnarray}
\noindent where $p' := \frac{dp}{d\tau}$.

\begin{figure}[!ht] 
    \includegraphics[width=10cm]{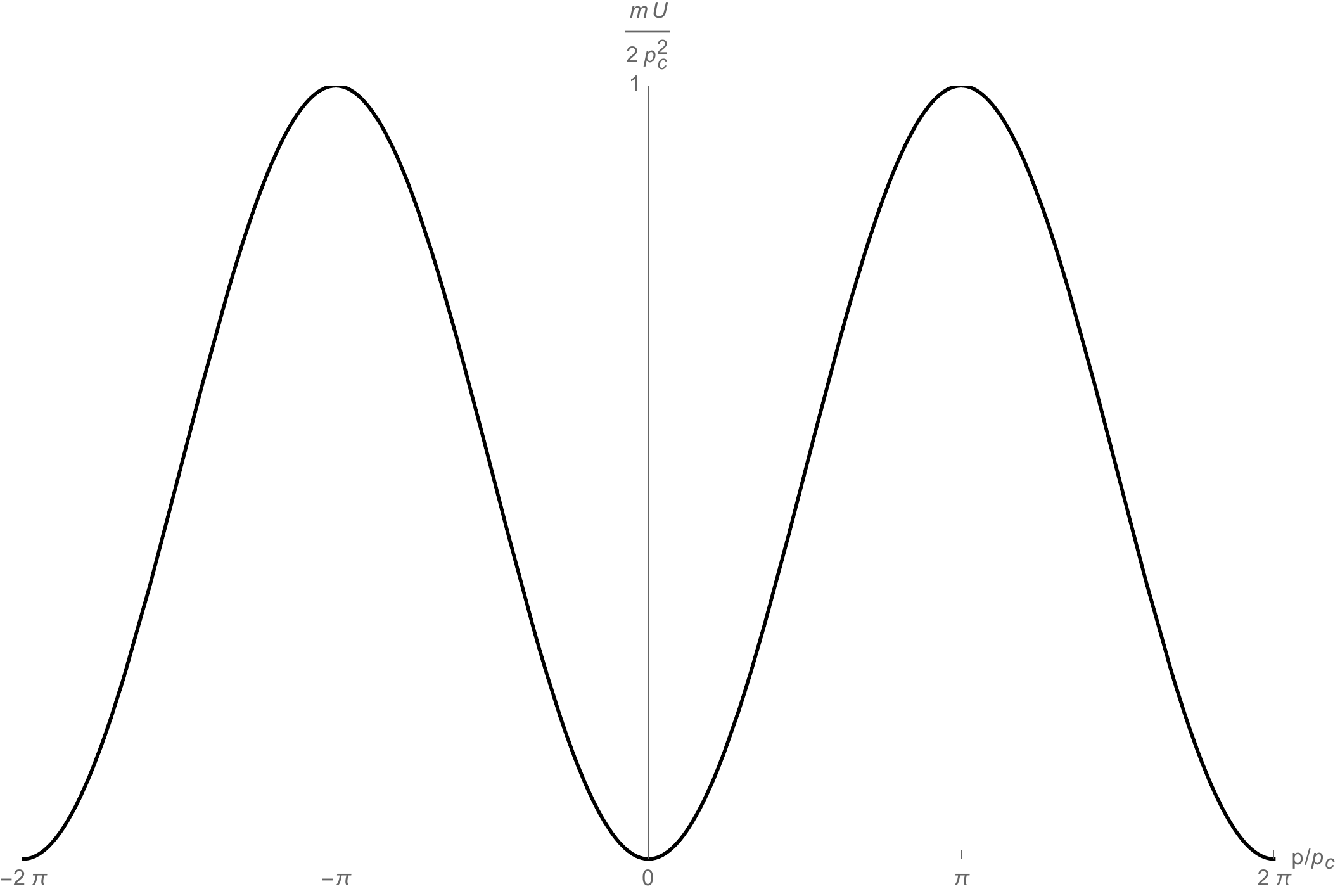} 
  \caption{Graph of the Potential $V(p)$ to notice the presence of the valleys allowing the existence of Instantons} \label{GraphPotential}
\end{figure}

As we already notice, the potential $V(p)$ is periodic with period $2 \pi p_c$ and it is null on the points $p= \{ \dots , - 2\pi p_c , 0, 2 \pi p_c , \dots \}$. See Figure \ref{GraphPotential}. Its domain comprises the entire real line as we are dealing with the amplitude without topological constraints given in (\ref{RegAmplitude}). Each of these points represents an edge in which the instanton solution can be evaluated for their corresponding value. The potential expanded around $p=0$ is of the form $V(p) \approx \frac{1}{2m} p^2$ and therefore, behaves as the kinetic energy term of the simple harmonic oscillator. This explains the selection of this system as a regulator in the previous section.

The Euler-Lagrange equation derived from (\ref{MLagrangian}) is given by
\begin{eqnarray}
 \frac{p'' }{k} - \frac{p_c}{m} \sin\left( \frac{p}{p_c} \right) = 0, \label{Pendulum}
\end{eqnarray}
\noindent and by defining the dimensionless variable $\xi :=  p/ p_c$ the previous equation takes the familiar classical pendulum equation in imaginary time \, $ \xi''=\omega^2 \sin(\xi) $. The integration of equation (\ref{Pendulum}) gives 
\begin{equation}
 p'^2 = 2k V(p) + K , \label{EC1}
\end{equation}
\noindent where $K$ is the first integration constant. A requisite for a solution to be an instanton type solution is to yield a finite value for the Euclidean action. This imposes the condition $K=0$ on the previous expression. Let us denote by a capital $P$ the instanton solution with equation of the form
\begin{equation}
 P' = \pm \sqrt{ 2k V(P)}. \label{PINst}
\end{equation}

Selecting the positive root and integrating the equation (\ref{PINst}) we obtain the instanton solution to be of the form
\begin{equation}
P^+ = 4 p_c \arctan \left[ e^{ \omega(\tau - \tau_c)}  \right] , \label{SolOP}
\end{equation}
\noindent where $\tau_c$ is the other integration constant named center of the instanton.

\begin{figure}[!ht] 
    \includegraphics[width=10cm]{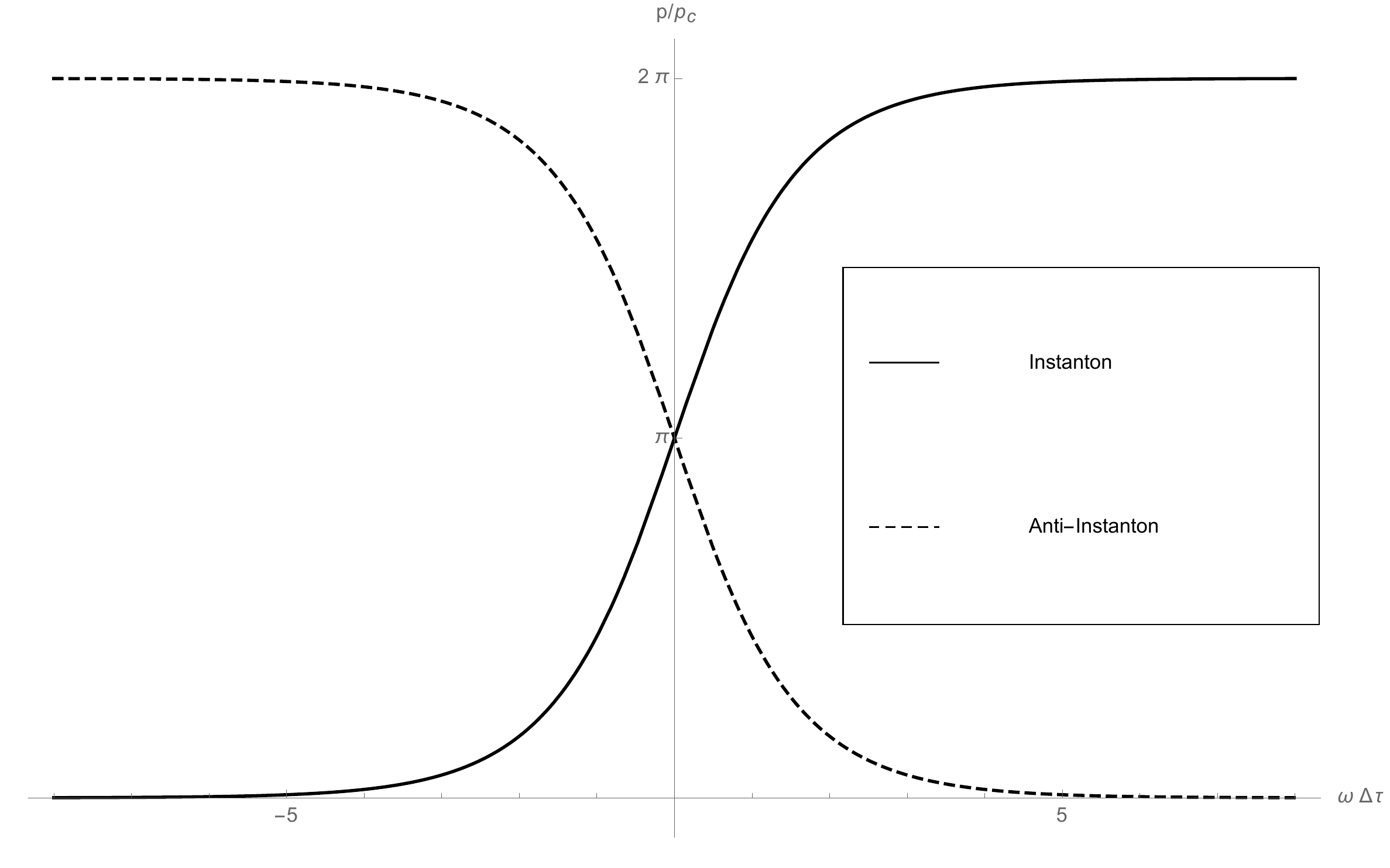} 
  \caption{Polymer Instanton and Polymer Anti-Instanton} \label{InstantonPolimerico}
\end{figure}

It is easy to check that this solution satisfies the equation (\ref{EC1}). We also observe that in the limit when $\Delta \tau:= \tau - \tau_c \rightarrow + \infty$ then $P^+ \rightarrow + 2 \pi p_c $ and in the limit $\Delta \tau \rightarrow - \infty$ then $P^+ \rightarrow 0$. These results are adequate in the spirit that on these limits (of large imaginary time) the instanton arrives at the edges of the interval (Figure \ref{InstantonPolimerico}), recalling that the potential $V(p)$ is such that $ V(p = 2 n \pi p_c) = 0$. The negative root in equation (\ref{PINst}) gives an anti-instanton solution. $P^+(\tau)$ and $P^{-}(\tau)$ contribute to the amplitude of the full quantum system when the other parts of the periodic potential are considered.

The potential $V(p)$ evaluated on this solution reads
\begin{equation}
V(P^+) = \frac{2 p^2_c}{ m} \mbox{Sech}^2(\omega \Delta \tau), \label{PolyInst1}
\end{equation}
\noindent and notice that in the large time intervals $\Delta \tau \rightarrow +\infty$ the potential $V(P^+) \rightarrow 0$, i.e., remains finite. This has the same effect on the action   
\begin{equation}
 S^{PHO}_E[P^+] = \int^{+\infty}_{-\infty} d\tau \, [ 2V(P^+)] = \frac{4 p^2_c}{ m \omega} \mbox{Tanh} (\omega \Delta \tau)|^{+\infty}_{-\infty} =  \frac{8 p^2_c}{ m \omega} , \label{IVAction}
\end{equation}
\noindent and confirms that $P^+$ is in rigor an instanton. Observe that the value of the action on this instanton tends to infinity when $\mu \rightarrow 0$. This means that in this limit, there is no instanton-like solution. For a fixed but small $\mu$, this finite value of the action evaluated on the solution $P^+$ is a direct consequence of the `instantonic' character of the solution.

 Let us move to the Euclidean Hamiltonian analysis of this solution. Consider the action (\ref{MLagrangian}) and let us define the coordinate variable in Euclidean time given by
\begin{equation}
x = \frac{\partial L}{\partial p'} = \frac{p'}{k}. \label{CInstanton}
\end{equation}
\noindent With this definition, the Euclidean Hamiltonian is given by $H := p' x - L$, hence, the Euclidean polymer Hamiltonian takes the form
\begin{equation}
H[x,p] =  \frac{k x^2}{2}  - V(p).
\end{equation}

The Hamiltonian evaluated on the instanton solution gives
\begin{equation}
H[X, P] =  \frac{k X^2}{2}  - V(P) = 0,
\end{equation}
\noindent which can be interpreted as an Euclidean classical trajectory with null energy.
 
\begin{figure}[!ht] 
    \includegraphics[width=8cm]{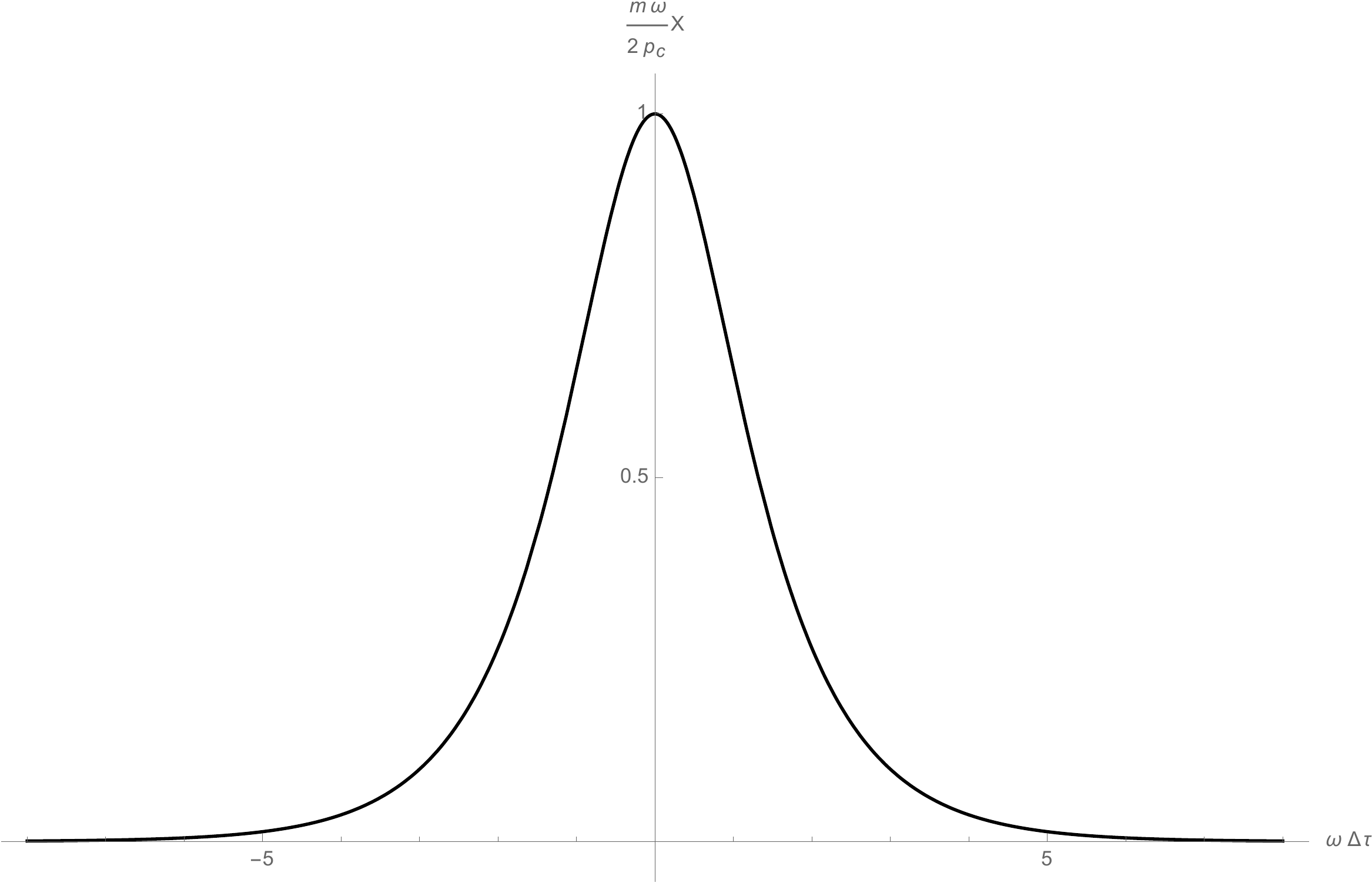} 
  \caption{Polymer Instanton in coordinates $X^+$} \label{XInstantonPolimerico}
\end{figure}

\begin{figure}[!ht] 
    \includegraphics[width=4cm]{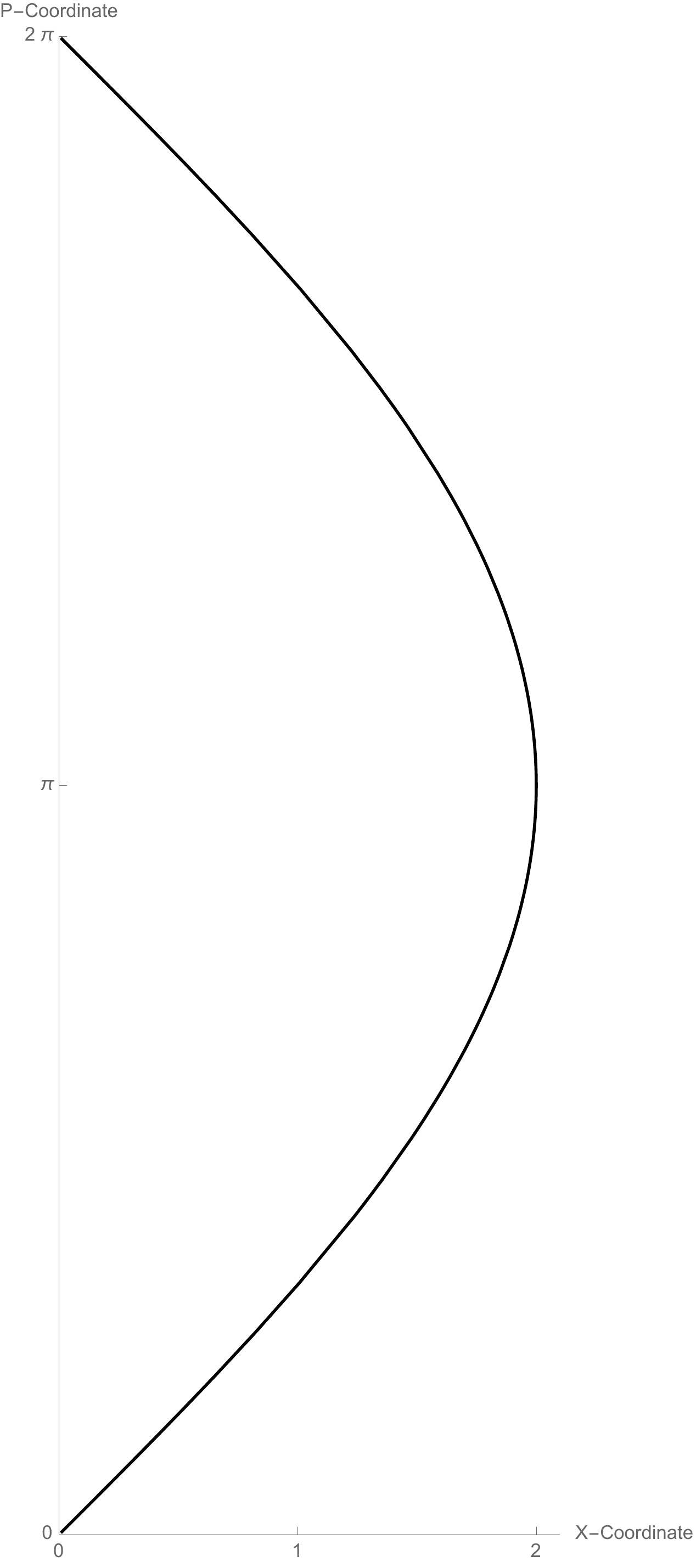} 
  \caption{Instanton in phase space with $m \omega =1$ and $p_c =1$} \label{InstantonPS}
\end{figure}

The explicit form of the instanton solution coordinate can be derived using (\ref{CInstanton})
\begin{equation}
X^+ = \frac{2 p_c}{m \omega} \mbox{Sech}\left( \omega \Delta \tau \right), 
\end{equation}
\noindent which peaked in the vicinity of $\Delta \tau =0$ as can be seen in Figure (\ref{XInstantonPolimerico}). For an infinite time interval, the coordinate tends to zero in accordance to the momentum behavior.
With these expressions we can draw an instanton path in the Euclidean phase space with cartesian coordinates
\begin{equation}
(X^+, P^+) = 2 p_c \left( \frac{1}{m \omega} \mbox{Sech}\left( \omega \Delta \tau \right),2 \tan^{-1} \left[ e^{ \omega \Delta \tau}  \right]  \right).
\end{equation}
In the Figure (\ref{InstantonPS}) we draw the instanton in the phase space with coordinates $(X^+(\Delta \tau), P^+(\Delta \tau))$. It can be seen that when $\Delta \tau \rightarrow -\infty$ the instanton starts at $(0, 0)$ and when $\Delta \rightarrow +\infty$ the instanton arrives at the point $(0, 2  \pi p_c)$.

Let us briefly study the instanton solution of the Euclidean action of the harmonic oscillator in momentum variables. In this case, the action takes the form
\begin{equation}
{\cal S}^{(H)}_E = \int^{\tau_f}_{\tau_i} d\tau\, \left[ \frac{1}{2k} p'^2 + \frac{1}{2m}p^2 \right],
\end{equation}
\noindent and the equation of motion is given by
\begin{equation}
p'' = \omega^2 p.
\end{equation}
 The general solution of this equation is 
\begin{equation}
P^{(H)} = c_0 e^{\omega \Delta \tau} + c_1 e^{-\omega \Delta \tau},
\end{equation} 
\noindent where $c_0$ and $c_1$ are arbitrary coefficients. The action ${\cal S}^{(H)}_E$ evaluated on $P^{(H)}$ in the finite time interval $(\tau_i,\tau_f)$ takes the following form
\begin{equation}
S^{H}_E[P^{(H)}] = \frac{1}{2m\omega} \left[ c^2_0 e^{2\omega \Delta \tau} + c^2_1 e^{-2\omega \Delta \tau} \right]^{\tau=\tau_f}_{\tau=\tau_i}.
\end{equation}
If we consider the limit $\tau_f, \tau_i \rightarrow +\infty, -\infty$ then only when $c_0=c_1=0$ the Euclidean action ${\cal S}^{H}_E[P^{(H)}]$ is finite with value $S^{H}_E[P^{(H)}] = 0$. This probes that in the case of the harmonic oscillator the only instanton solution is the trivial solution, i.e., $P^{(H)} = 0$. We will see further that although the instanton solution is $P^{(H)}=0$, the quantum contributions to the action $S^{H}_E[P^{(H)}]$ around the $P^{(H)}$ are non-trivial and yields the appropriate regularization of the quotient in (\ref{RegAmplitude}). 

Summarizing this section, we have obtained an instanton solution $P^+$ given in (\ref{SolOP}) for the Euclidean action (\ref{MLagrangian}). The Euclidean action evaluated on this solution gives (\ref{IVAction}) and the instanton fulfills the conditions $P^+(+\infty) = 2 \pi p_c$ and $P^+(-\infty) =0$. For the harmonic oscillator we found that the instanton solution is $P^{(H)}=0$ and its action takes the value $S^{H}_E[P^{(H)}] = 0$. With these results, we are ready to move to the next section and study the quantum fluctuation around these instantons solutions.


\section{Quantum instanton fluctuation and penetration barrier amplitude} \label{QFInst}

With the results of the previous section let us return to the amplitude (\ref{RegAmplitude}) and let us consider a Wick rotation $t \rightarrow i \tau$. Now let us expand each Euclidean action in this amplitude up to second order around their corresponding instanton solution $P^+$ and $P^{(H)}$. The deviations of the trajectories are given as $\delta p = p - P^+$ for the Euclidean action ${{\cal S}_E}$ and $\delta p^H $ for ${\cal S}^{(H)}_E$. The amplitude (\ref{RegAmplitude}) is now written as
\begin{eqnarray}
\langle \langle p_f, \tau_f | p_i, \tau_i \rangle \rangle &=& \langle p_f, \tau_f | p_i, \tau_i \rangle^{(H)}  \frac{ e^{- \frac{1}{\hbar} S^{PHO}_E[P^+] } \int  {\cal D} \delta p \,  e^{ -\frac{1}{ \hbar} \int^{\tau_f}_{\tau_i} d\tau \left\{ \frac{\delta p}{2} \left[ - \frac{1}{k} \frac{d^2}{d\tau^2} + V''(P^+) \right] \delta p+ {\cal O}(\delta p^3) \right\}} }{e^{- \frac{1}{\hbar} S^{H}_E[P^H] } \int  {\cal D} \delta p^H \,  e^{ -\frac{1}{ \hbar} \int^{\tau_f}_{\tau_i} d\tau \,\frac{\delta p^H}{2} \left[ -\frac{1}{k} \frac{d^2}{d\tau^2} + \frac{1}{m} \right]\delta p^H } } . \label{AMPCla}
\end{eqnarray}
\noindent where the potential $V''(P^+) :=  \left(\frac{\partial^2 V(p)}{\partial p^2}\right)_{p=P^+} $ is given by
\begin{equation}
 V''(P^+) =  \frac{1}{m} \left[ 1- 2 \, \mbox{Sech}^2( \omega \Delta \tau) \right]. \label{PotentialPHO}
 \end{equation}
 The zeroth order in this expansion gives the classical values of the Euclidean action on the instanton solutions whose values were determined in the previous section.
 
 Discarding the third order terms $ {\cal O}(\delta p^3)$, the remaining integrals are Gaussian-type integrals in the variables $\delta p$ and $\delta p^H$. To solve them, we first propose that the operators $- \frac{1}{k} \frac{d^2}{d\tau^2} + V''(P^+)$ and $- \frac{1}{k} \frac{d^2}{d\tau^2} + \frac{1}{m}$ can be diagonalized \cite{Coleman, Rajamaran, ABCInstanton, Kazama}. Their corresponding eigenvalues equations are 
\begin{equation}
 \left( -\frac{1}{k} \frac{d^2}{d\tau^2} + V''[P^+] \right) f_n(\tau) = \epsilon^{PHO}_n f_n(\tau), \label{QIEquation}
\end{equation}
\noindent for the polymer harmonic oscillator and
\begin{equation}
\left( - \frac{1}{k} \frac{d^2}{d\tau^2}  + \frac{1}{m} \right) f^H_n = \epsilon^{H}_n f^H_n, \label{QIEquationHO}
\end{equation}
\noindent for the harmonic oscillator. The eigenfunctions $f_n(\tau)$ and $f^H_n(\tau)$ are related with the deviations $\delta p$ and $\delta p^H$ as
\begin{equation}
\delta p = \sum^{+\infty}_{n=0} c_n \, f_n(\tau), \qquad  \delta p^H = \sum^{+\infty}_{n=0} c^H_n \, f^H_n(\tau),
\end{equation}
\noindent and we assume in this notation that the spectrum is discrete. Additionally, we require both systems of eigenfunctions $f_n$ and $f^H_n$ to be orthonormal in the integration interval $(\tau_i,\tau_f)$. Using these definitions, the integrations in the variables $c_n$ and $c^H_n$ in the amplitude (\ref{AMPCla}) give the following expression for the amplitude 
\begin{eqnarray}
\langle \langle p_f, \tau_f | p_i, \tau_i \rangle \rangle&=& \langle p_f, \tau_f | p_i, \tau_i \rangle^{(H)}  \frac{ e^{- \frac{1}{\hbar} S^{PHO}_E[P^+] } }{ e^{- \frac{1}{\hbar} S^{H}_E[P^H] } } \prod^{+\infty}_{n=0} \left( \frac{\epsilon^{PHO}_n}{\epsilon^{H}_n} \right)^{- \frac{1}{2}}. \label{PAmpli}
\end{eqnarray}
 
The eigenvalue problem of the equation (\ref{QIEquationHO}) is easily solved. It is a Schr\"odinger-type equation for a particle in a constant potential. In order to obtain a discrete spectrum, we impose the `temporal box' boundary conditions, which in this case reads as $f^H_n(\tau_f) = f^H_n(\tau_i) = 0$. The `length' of the box is $ \tau_0 := \tau_f- \tau_i$ and the eigenvalues are
\begin{equation}
\epsilon^{H}_n = \frac{1}{m} + \frac{n^2 \pi^2}{m \omega^2 \tau^2_0}, \qquad n=0,1,2,\dots
\end{equation}

\begin{figure}[!ht] 
    \includegraphics[width=10cm]{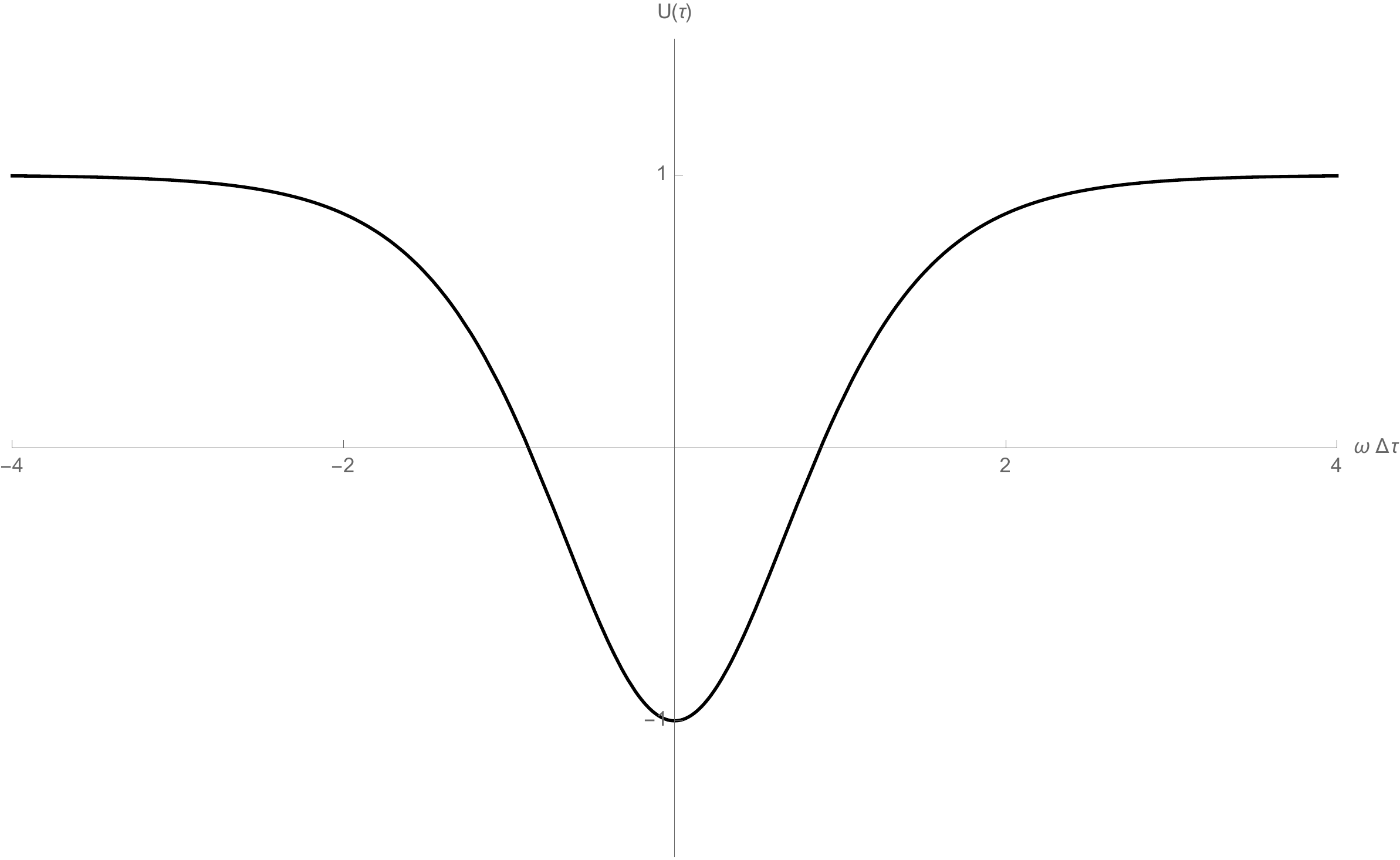} 
  \caption{Potential $U(\tau)$ associated to the quantum fluctuation $\delta p$} \label{GPotencialQ}
\end{figure}

The solution of the eigenvalue problem (\ref{QIEquation}) is a bit more complicated due to the potential (\ref{PotentialPHO}). Equation (\ref{QIEquation}) can be written into a more familiar form by giving it a Schr\"odinger type form
\begin{equation}
\frac{d^2}{d\tau^2}f_n(\tau) + \omega^2 \left[ m \epsilon - U(\tau) \right] f_n(\tau) = 0 , \label{PolyInst}
\end{equation} 
\noindent where the quantum potential $U(\tau)$ reads as
\begin{equation}
U(\tau) = 1 - 2 {\mbox{Sech}^2(\omega \Delta \tau)} , \label{FQPot}
\end{equation}
\noindent and its graph is given in Figure \ref{GPotencialQ}. 

Remarkably, the form of this potential for the quantum fluctuation is similar to that appearing in the system with the double-well potential $\tilde{V}(q) \sim (q^2 - \alpha^2)^2 $ \cite{ABCInstanton, Kazama}. In this case, the potential of the quantum fluctuation is $\tilde{U}(\tau) = 1 - \frac{3}{2} {\mbox{Sech}^2(\frac{1}{2} \omega \Delta \tau)}$. As can be seen, it changes by two numerical factors but, as we will see further, this small change is sufficient to modify the discrete spectrum. 

The spectrum for the potential $U(\tau)$ will be discrete if $\epsilon^{PHO}$ is such that $- \frac{1}{m} < \epsilon^{PHO} < +\frac{1}{m}$ and will be continuous if $\epsilon^{PHO} > + \frac{1}{m}$. In the Appendix \ref{SQFEq} we summarize the calculation of the solution to the equation (\ref{QIEquation}) together with the analysis of its eigenvalues. 

We obtain that the spectrum in the case of the equation (\ref{QIEquation}) is $\epsilon^{PHO}= \{ \epsilon^d_n, \epsilon^c \}$. Where the discrete part $\epsilon^d_n $ contains only the zeroth mode contribution $\epsilon^d_0 = 0 $ and the continuous part $\epsilon^c$ is given by the expression (\ref{QFMomentum}) and is labelled by a continuous variable $\tilde{p}$. The continuous part of the spectrum, $\epsilon^c$, becomes discrete after imposing the `temporal box' boundary conditions into its eigenfunctions and the relation (\ref{QFMomentum}) can be written as
\begin{equation}
\epsilon^c_n = \frac{1}{m} + \frac{(n \pi + \delta_{\tilde{p}})^2}{m (\omega \tau_0)^2} , \quad n=0,1,2,\dots\label{CSpectrum}
\end{equation}
\noindent where $ \delta_{\tilde{p}}$ is the relative phase of the eigenfunctions at infinity imaginary time.

The zeroth mode $\epsilon^d_0 =0$ must be treated apart because the integration of the variable $dc_0$ diverges in the interval $\tau \in (- \infty, + \infty)$. The standard treatment of this mode requires a relation between the variable $c_0$ and the center of the instanton $\tau_c$ \cite{ABCInstanton, Kazama}. In our case, the relation is given by 
\begin{equation}
\frac{dc_0}{\sqrt{2\pi}} = \frac{1}{f_0(\tau)} \left( \frac{dP^+}{d\tau}\right) \frac{d\tau_c}{\sqrt{2\pi}} = \sqrt{ \frac{S^{PHO}_E[P^+]}{2\pi\, \hbar} } \omega\, d\tau_c, \label{CPInt}
\end{equation}
\noindent where the expression (\ref{INSTZeroth}) of the appendix (\ref{SQFEq}) was used. 


Once we have the spectrum of both equations (\ref{QIEquation}) and (\ref{QIEquationHO}) we are ready to determine the penetration amplitude given in (\ref{PAmpli}).  First of all, recall that within the interval $[- \pi p_c, + \pi p_c)$ the quantum potential of the polymer harmonic oscillator is null at the point $p =0$. This implies that the penetration amplitude given in (\ref{PolyAmplitudeSum}) must be calculated as
\begin{equation}
\langle 0 , + \frac{\tau_0}{2} | 0 , - \frac{\tau_0}{2} \rangle = \sum_{n \in \mathbb{Z}} e^{2 \pi i n \frac{\lambda}{\mu}}  \langle \langle 2 \pi n p_c , + \frac{\tau_0}{2} | 0 , - \frac{\tau_0}{2} \rangle \rangle, \label{FAmplitude}
\end{equation}
\noindent hence, in order to determine $\langle \langle 2 \pi n p_c , + \frac{\tau_0}{2} | 0 , - \frac{\tau_0}{2} \rangle \rangle $, let us we first consider $n=1$. In this case (\ref{PAmpli}) takes the form
\begin{equation}  \label{PieceAm}
\langle \langle 2 \pi p_c , + \frac{\tau_0}{2} | 0 , - \frac{\tau_0}{2} \rangle \rangle =  \langle 0, + \frac{\tau_0}{2} | 0, - \frac{\tau_0}{2} \rangle^{(H)}  e^{- \frac{1}{\hbar} S^{PHO}_E[P^+]  } \frac{dc_0}{\sqrt{2\pi}}\prod^{+\infty}_{n=0} \left( \frac{\epsilon^{c}_n}{\epsilon^{H}_n} \right)^{- \frac{1}{2}},
\end{equation}
\noindent and notice that for the harmonic oscillator amplitude, the initial and final points are the same because the potential has only one minimum. 

Let us denote by $R$ the product of the ratios of the eigenvalues $\epsilon^{c}_n$ and $\epsilon^H_n$ and notice it can be written in terms of the momenta $\tilde{p}_n$ and $p_n$ as
\begin{eqnarray}
R &=& \prod^{+\infty}_{n=0} \frac{\epsilon^{PHO}_n}{\epsilon^{H}_n} =  \prod^{+\infty}_{n=0} \frac{\omega^2 + \tilde{p}^2_n}{\omega^2 + p^2_n} , 
\end{eqnarray}
\noindent where $\tilde{p}_n := \omega \sqrt{m \epsilon^c_n -1} = (n\pi + \delta_{\tilde{p}})/\tau_0$ and $p_n := \omega \sqrt{m \epsilon^{H}_n -1} = n\pi/\tau_0$. The difference in these momenta for high values of $\tau_0$ is approximately 
\begin{equation}
\overline{\Delta} p_n := \tilde{p}_n - p_n = \delta_{\tilde{p}} / \tau_0 \approx \frac{\delta_{\tilde{p}}}{\pi} \Delta p_n,
\end{equation}
\noindent where in the last relation we inserted $\Delta p_n = p_{n+1} - p_n = \pi/\tau_0$. The expression for $R$ using this approximation is given by
\begin{eqnarray}
R &=& \prod^{+\infty}_{n=0} \frac{\omega^2 + \tilde{p}^2_n}{\omega^2 + p^2_n} \approx \exp\left\{ \sum^{+\infty}_{n=0} \frac{2 p_n \overline{\Delta}p_n }{\omega^2 + p^2_n}   \right\} \approx \exp\left\{ - \frac{1}{\pi} \int^{+\infty}_0  dp \, \ln\left( 1 + \frac{p^2}{\omega^2} \right) \frac{d}{dp} \delta  \right\} = \frac{1}{4}. \label{CPSpectrumQ}
\end{eqnarray}
Let us now integrate the continuous part (\ref{CPInt})
\begin{equation}
\int^{+\tau_0/2}_{-\tau_0/2} \, \sqrt{ \frac{S^{PHO}_E[P^+]}{2\pi\, \hbar} } \omega\, d\tau_c = \sqrt{ \frac{S^{PHO}_E[P^+]}{2\pi\, \hbar} } \omega\tau_0,
\end{equation}
\noindent and together with the result for $R$, let us insert them in the amplitude (\ref{PieceAm}) 
\begin{eqnarray}
\langle \langle 2 \pi n p_c , + \frac{\tau_0}{2} | 0 , - \frac{\tau_0}{2} \rangle \rangle = \langle 0, + \frac{\tau_0}{2} | 0, - \frac{\tau_0}{2} \rangle^{(H)}  \rho(\tau_0), 
\end{eqnarray}
\noindent where $\rho(\tau_0)$ is the `density of instantons' \cite{ABCInstanton, Kazama} 
\begin{equation}
\rho(\tau_0) = \sqrt{\frac{2 S^{PHO}_E[P^+]}{\pi \, \hbar} } \, \omega \tau_0\,e^{- \frac{S_E[P]}{\hbar}}. \label{Rhozero}
\end{equation}

We now move to the  long time regime which allow us the introduction of multiple instanton and anti-instanton solutions. This model is called `dilute instanton gas approximation'. In this scenario, the time $\tau_0$ is large enough to allow widely separated pseudo-particles (instantons and anti-instantons) fulfilling the boundary conditions \cite{Coleman, Rajamaran, ABCInstanton, Kazama}. This feature leads to the band structure of the spectrum for the case of periodic potentials \cite{BSimon2}. In this context, let us consider the amplitude of a `dilute instanton gas' contribution from $p_i=0$ to the point $p_f = 2 \pi n p_c$
\begin{eqnarray}
\langle \langle 2 \pi n p_c , + \frac{\tau_0}{2} | 0 , - \frac{\tau_0}{2} \rangle \rangle &=&  \langle 0, + \frac{\tau_0}{2} | 0, - \frac{\tau_0}{2} \rangle^{(H)} \sum_{n_1, n_2 =0} \frac{1}{n_1 ! n_2 !} \rho(\tau_0)^{n_1 + n_2} \delta_{n_1 - n_2, n}, \nonumber \\
&=&  \langle 0, + \frac{\tau_0}{2} | 0, - \frac{\tau_0}{2} \rangle^{(H)}
\int^{2 \pi}_0 \frac{d \theta}{2 \pi} e^{i n \theta} e^{ 2 \rho(\tau_0) \cos \theta }, \nonumber \\
&=& \langle 0, + \frac{\tau_0}{2} | 0, - \frac{\tau_0}{2} \rangle^{(H)} i^n J_n\left( -i 2 \rho(\tau_0) \right), \label{MICont}
\end{eqnarray}
\noindent where $J_n(x)$ is the Bessel function of the first kind \cite{Abramovitz}. Let us insert (\ref{MICont}) into (\ref{FAmplitude}) and make explicit the dependence in the parameter $\lambda$
\begin{eqnarray}
\langle 0 , + \frac{\tau_0}{2} | 0 , - \frac{\tau_0}{2} \rangle^{(\lambda)} &=& \langle 0, + \frac{\tau_0}{2} | 0, - \frac{\tau_0}{2} \rangle^{(H)}  \sum_{n \in \mathbb{Z}}e^{ 2 \pi i n \frac{\lambda}{\mu}}   i^n J_n\left( -i 2 \rho(\tau_0) \right) , \nonumber \\
&=& \langle 0, + \frac{\tau_0}{2} | 0, - \frac{\tau_0}{2} \rangle^{(H)}  \exp\{ - i 2 \rho(\tau_0) \left[ i e^{2 \pi i \frac{\lambda}{\mu}} + i e^{- 2 \pi i \frac{\lambda}{\mu}} \right] \}, \nonumber \\
& =& \langle 0, + \frac{\tau_0}{2} | 0, - \frac{\tau_0}{2} \rangle^{(H)}  e^{ 2 \rho(\tau_0) \cos(2 \pi \frac{\lambda}{\mu})   } , \label{FAmpPHO}
\end{eqnarray}
\noindent where in the second line we use the property of the Bessel functions $$ e^{\frac{x}{2} (t - \frac{1}{t})} = \sum_{n\in \mathbb{Z}} t^n J_n(x). $$

Finally, recall that in the long time regime ($\tau_0 \rightarrow + \infty$), the Euclidean amplitude of the harmonic oscillator can be approximated as
\begin{equation}
\langle 0, + \frac{\tau_0}{2} | 0, - \frac{\tau_0}{2} \rangle^{(H)} \sim | \Psi_0(0)|^2 e^{- \frac{\tau_0}{\hbar} E_0},
\end{equation}
\noindent where $\Psi_0(0)$ is the vacuum eigenstate of the system at $p=0$ and $E_0$ its eigenvalue. Combining this result with (\ref{FAmpPHO}) we obtain that the main contribution to the amplitude in the long time regime takes the form
\begin{equation}
\langle 0 , + \frac{\tau_0}{2} | 0 , - \frac{\tau_0}{2} \rangle^{(\lambda)} \approx  | \Psi_0(0)|^2 e^{- \frac{\tau_0}{\hbar} \left[ E_0 -  \frac{2\hbar}{\tau_0} \rho(\tau_0) \cos(2 \pi \frac{\lambda}{\mu}) \right] }.
\end{equation}

The energy inside the square brackets in the previous expression gives the energy of the system
\begin{equation}
E^{\lambda}_0 = \frac{\hbar \omega}{2} \left[ 1 - \frac{4\, l_0}{\mu\, \sqrt{\pi}} \cos\left( \frac{2 \pi \lambda}{\mu}\right)  e^{- 8 \left(\frac{l_0}{ \mu}\right)^2 } \right], \label{InstEigenvalue}
\end{equation}
\noindent  where $l_0 := \sqrt{\frac{\hbar}{m \omega}}$ is the characteristic length of the vacuum wavefunction of the standard harmonic oscillator. Each eigenvalue $E^\lambda_0$, with $\lambda \in [0,\mu)$, is an approximation, in the long time regime, of the zeroth eigenvalue of the Hamiltonian (\ref{PHamiltonian}) in the Hilbert space ${\cal H}^{(\lambda)}_{poly}$. 

The first point to be notice in the expression (\ref{InstEigenvalue}) is that it gives a simple and compact  expression for the the lowest energy eigenvalues of the polymer harmonic oscillator. In order to compare the accuracy of this result, we graph the zeroth eigenvalue for the periodic Mathieu equation, i.e., without using the instanton approximation and $E^{\lambda = 0}_0$ given in (\ref{InstEigenvalue}) in terms of $\mu/l_0$
\begin{figure}[!ht] 
    \includegraphics[width=10cm]{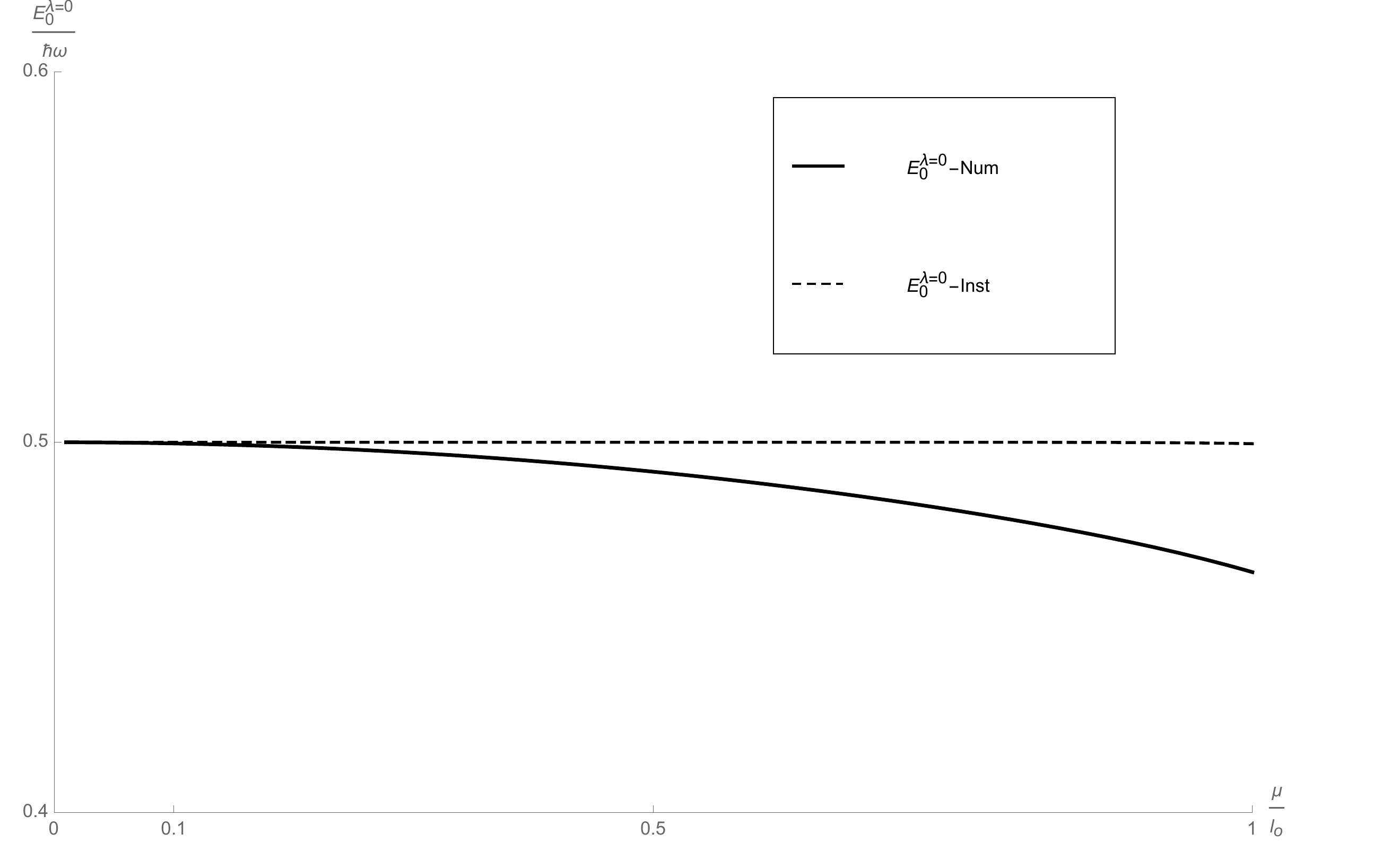} 
  \caption{Zeroth eigenvalue using instanton methods and the exact (numerical) calculation.} \label{Comparacion}
\end{figure}

 It can be seen in Figure (\ref{Comparacion}), that when $\mu/ l_0 \sim 10^{-1}$ the eigenvalue $E^{\lambda=0}_0$ of the instanton description coincides with the exact numerical value $E_0$-Num of the periodic Mathieu solution. Therefore, our approximation is valid in the interval $ \mu/l_0 \lesssim 10^{-1} $. On the other hand, recall that in order to discard polymer effects (or spatial discreteness) on the quantum harmonic oscillator, then $\mu \lesssim 10^{-19} \, m$. If we consider the standard textbooks values for the harmonic oscillator parameters then $l_0 \sim 10^{-12} \, m$. Combining these values we obtain the condition $\mu /l_0 \lesssim 10^{-7}$ \cite{ShadowsStates}. Consequently, the instanton analysis can be very well used to described the physics of the polymer harmonic oscillator. 

\begin{figure}[!ht] 
    \includegraphics[width=10cm]{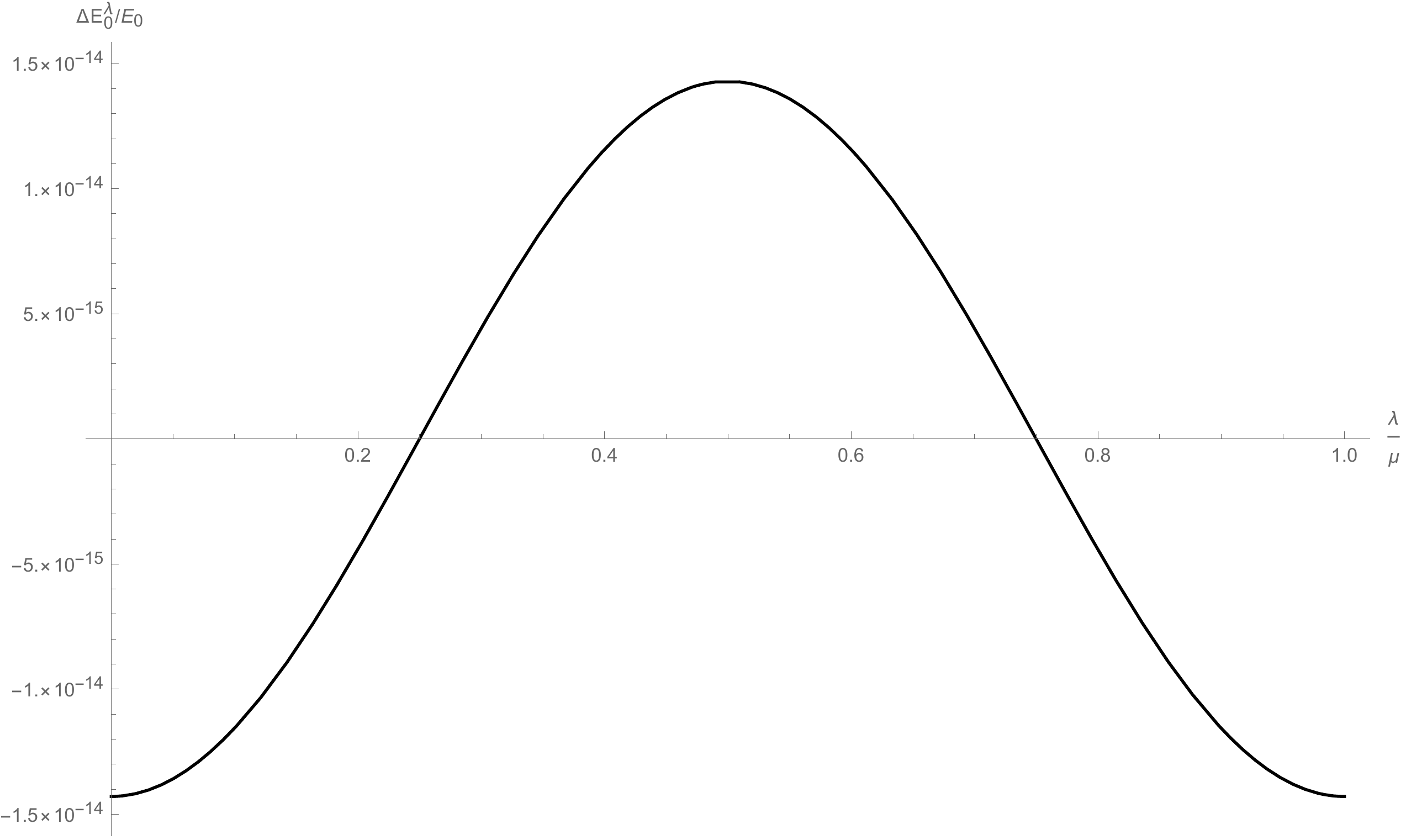} 
  \caption{$\Delta E^\lambda_0/ E_0$ as function of the quotient $\lambda/\mu$ with $\mu = 2 l_0$.} \label{Gap}
\end{figure}

A remarkable aspect of the expression (\ref{InstEigenvalue}) is that it shows the band structure mentioned by Barbero et al. \cite{Barbero}. The minimum eigenvalue of the band is the one corresponding to $\lambda =0$ while the supremum is the one corresponding to $\lambda = 0.5 \mu$ (see Figure (\ref{Gap})).
 Any other eigenvalue in the band with different $\lambda$ is doubly degenerate (recall that $\lambda = \mu$ is not allowed) and their eigenstates are in the Hilbert spaces given by $\lambda$ and $\mu - \lambda$.  These results are in complete agreement with the theory of periodic potentials given in \cite{ReedSimon} and with the spectral analysis of almost periodic Schr\"odinger operators \cite{Chojnacki}. This double degeneracy of the spectrum (for $\lambda \neq 0$ and $\lambda = \mu/2$) implies that any given state $| \Psi_0(t) \rangle$ within the first energy band can be written as 
\begin{equation}
|\Psi_0(t) \rangle = \oplus_{\lambda \in [0,\mu/2)} c^{(\lambda)} e^{- i\frac{t}{\hbar} E^{\lambda}_0}| 0, \lambda \rangle, 
\end{equation}
\noindent where $| 0, \lambda \rangle $ is the eigenstate in ${\cal H}^{(\lambda)}_{poly}$ corresponding to the eigenvalue $E^{\lambda}_0$ and the arbitrary constants $c^{(\lambda)}$ are non-zero only at a countable set of $\lambda$ values. The width of this energy band, named $E^{w}_0$, is given by
\begin{equation}
E^{w}_0 := E^{\lambda = 0.5 \mu}_0 - E^{\lambda =0}_0 = \frac{4 \hbar \omega}{\sqrt{\pi}} \frac{l_0}{\mu}e^{- 8 \left(\frac{l_0}{\mu}\right)^2}, \label{BandStructure}
\end{equation}
\noindent and if we consider $\mu/l_0 \sim 10^{-7}$, then the width is a very small quantity $ E^{w}_0 \sim  \frac{\hbar \omega}{2} 10^{-10^{14}} eV$. A photon with this energy $E^{w}_0$, emitted by a polymer oscillator with frequency $\omega \sim 10^{15} s^{-1}$ has a wavelength which is a million times larger than the  diameter of the visible universe. In other words, if we consider $\mu \sim 10^{-19}\, m $ then the deviation of the $E^{\lambda}_0$ from $E_0$ is not experimentally tested.

Let us conclude the analysis of the expression (\ref{InstEigenvalue}) by considering the formal limit $\frac{\mu}{l_0} \rightarrow 0$. In this limit, all the energy eigenvalues take the form
\begin{equation}
\lim_{\frac{\mu}{l_0} \rightarrow 0} E^\lambda_0 = \frac{\hbar \omega}{2}, \label{Limit}
\end{equation}
\noindent  i.e., the first energy band `gets compressed' to yield the single vacuum eigenvalue $E_0 :=  \hbar \omega /2$ of the standard quantum harmonic oscillator. Thus, in this limit, the eigenvalue $E^\lambda_0 = \hbar \omega/2$ of the standard quantum harmonic oscillator can be seen as a degenerate eigenvalue and its uncountable degeneracy is labelled by $\lambda \in[0,\mu)$, as was pointed in \cite{Barbero}. Notice, however, that this degeneracy is only apparent: this limit is a mathematical trick and is used to provide a link between the standard quantization of the harmonic oscillator and its polymer version.

\section{Discussion} \label{Discussion}

Polymer models allow us to gain understanding of some of the techniques used in the loop quantization program. In the case of known quantum systems, it is crucial to recover their experimental results. This fact can be observed within the formal limit $\mu/l_0 \rightarrow 0$ on the polymer quantum harmonic oscillator. The parameter $\mu$ was introduced via the Hamiltonian operator $\widehat{H}^{(\mu)}_{poly}$. Such a Hamiltonian allows the splitting of the polymer Hilbert space ${\cal H}_{poly} = \int^{\oplus}_{\lambda \in [0,\mu)} {\cal H}^{(\lambda)}_{poly} d\lambda^c$ where ${\cal H}^{(\lambda)}_{poly} := L^2([-\pi \frac{\hbar}{\mu}, + \pi \frac{\hbar}{\mu}))$ and $d\lambda^c$ is a countable measure on the set $[0,\mu)$. The referred formal limit is usually taken in the super selected Hilbert space with $\lambda =0$. The analysis in the full polymer Hilbert space was carried out by Barbero et al. \cite{Barbero}. Of particular relevance on Barbero's derivation is the pure point spectrum nature of the polymer Hamiltonian,  $\sigma_{pp}({\widehat{H}}) = \cup_{n=0} [E^L_n , E^R_n]$, as a result of the non-regular representation. This is a typical feature of non-regular representations of almost periodic operators \cite{TheoremSH} and it is connected to the non-separability of the polymer Hilbert space.  As was pointed out in \cite{Barbero}, this feature of the Hilbert space renders difficult the analysis of the statistical mechanics of such systems. Essentially, the cardinality of the spectrum of the Hamiltonian operator $\widehat{H}^{(\mu)}_{poly}$ in the entire Hilbert space ${\cal H}_{poly}$ is $\#\mathbb{R}$. This implies the partition function $Z(\beta)$ to be infinity and therefore, the thermal density matrix $\rho=\frac{e^{-\beta \widehat{H}}}{Z}$ is ill-defined.

The pure point spectrum of the Hamiltonian is present in the band structure which appears as a consequence of the periodicity of the polymer Hamiltonian. In standard quantum mechanics for periodic potentials the bands correspond to the continuum spectrum and can be studied as a tunneling effect carried out by pseudo-particles named instantons. Motivated by this, and inspired by \cite{Barbero}, the purpose of this work was to establish a connection between the standard results of quantum mechanics for periodic potentials and lattice quantum mechanics, with those of polymer quantum mechanics. Particularly, we payed attention to the instanton methods in order to obtain similar and additional conclusions to those in \cite{Barbero} although by different ways.

To accomplish our task, we first calculated the renormalized propagator of the polymer harmonic oscillator. The `superselected nature' of the polymer Hilbert spaces ${\cal H}^{(\lambda)}_{poly}$ allows us to treat the propagator on each of the Hilbert spaces separately. The momentum variables within these ${\cal H}^{(\lambda)}_{poly}$ are topologically constrained. The techniques developed in \cite{Kleinert} were used to achieve the result (\ref{PolyAmplitudeSum}) together with (\ref{RegAmplitude}). The semiclassical potential (\ref{PolyPot}) yields nonlinear equations of motion which brings into consideration, the applicability of the instanton methods. 

Instantons in quantum mechanics are solutions of the Hamilton equations in imaginary time. $P^+$ given in (\ref{SolOP}) is the instanton of the polymer harmonic oscillator and it renders the finite value of the Euclidean action, given in (\ref{IVAction}). The quantum fluctuation around $P^+$ is used in order to obtain the quotient of the amplitudes appearing in the renormalized propagator (\ref{AMPCla}). The dynamic of such fluctuations is ruled by the Schr\"odinger type equation (\ref{QIEquation}) or by its simplified version (\ref{PolyInst}). The final amplitude is given in (\ref{FAmpPHO}).

The vacuum energy for long (imaginary) times can be derived from (\ref{FAmpPHO}) as is given by (\ref{InstEigenvalue}). Notably, vacuum energy depends on $\lambda$ due to our calculations were done for a fixed value of $\lambda$. That is to say, we derived the energy eigenvalues within the first allowed energy band of the polymer harmonic oscillator. Naturally, with this result we obtain the width of the first allowed band as can be noticed in (\ref{BandStructure}).

There are some worth mentioning aspects of this band structure that we have obtained. First, it shows directly the point spectrum that was mentioned in Barbero's work \cite{Barbero}. Point spectrum here is referred to the fact that the parameter $\lambda \in [0, \mu) \subset \mathbb{R}_c$.

 Secondly, in the limit $\frac{\mu}{l_0} \rightarrow 0$ the band width gives rise to the zeroth energy eigenvalue of the standard quantum harmonic oscillator as can be seen in (\ref{Limit}). This is a particularly interesting outcome which implies that the gap function in Figure (\ref{Gap}) `contracts' to a point. As a result, emerges an apparent uncountable infinite degeneracy of the eigenvalue $E_0$. If this analysis is expanded to the other bands, then it will imply that the effective degeneracy of each eigenvalue of the standard harmonic oscillator is again uncountable infinite as was already pointed out in \cite{Barbero}. When the limit is considered in the polymer amplitude, then the Green's function of the standard quantum harmonic oscillator is recovered. This result is independent of the parameter $\lambda$, which is a desired result.

In the third place, the band structure is fully consistent with the Floquet theory of quantum periodic potentials \cite{ReedSimon, BSimon, BSimon2}. A key point in this aspect is the adequate degeneracy of the zeroth eigenvalues \cite{Chojnacki} of the polymer harmonic oscillator. The lowest eigenvalue corresponds to $\lambda =0$ and the highest to $\lambda = \mu/2$. The other eigenvalues are doubly degenerated due to $\lambda \neq \mu$.

We compared the zeroth eigenvalue with the exact (numerical) solution of the periodic Mathieu equation. When $\mu/l_0 \sim 10^{-1}$, then the eigenvalue obtained with the instanton methods fits the exact (numerical) solution. Now, recall that, as we mentioned in the introduction, when the parameter $\mu$ is much more smaller than the characteristic length of the standard quantum harmonic oscillator $l_0$, the mean values of the polymer version of the observables cannot be separated of the mean values of the observables within the standard quantization. This will occur only in case that $\mu/l_0 \lesssim 10^{-7}$ \cite{ShadowsStates}. This implies that the instanton methods offers a description of the polymer harmonic oscillator which can be used within the interval $ 10^{-7} < \mu/l_0 \lesssim 10^{-1}$. Of course, when $\mu/l_0 \lesssim 10^{-7}$, polymer description is no longer required: the standard harmonic oscillator should be used. When $\mu/l_0 > 10^{-1}$, then instanton methods fail. It would be interesting to understand if these tools can be applied to quantum cosmology. Particularly, due to the possibility to derive simple and compact expressions analog to (\ref{InstEigenvalue}) for the energy eigenvalues or other physical quantities. 

 Finally, recall that if we are attending a process which is particular to a given super-selected Hilbert space, then it is suffice to use lattice quantum mechanics but, if on the other hand, our interest requires the dynamics on the full polymer Hilbert space ${\cal H}_{poly}$, then lattice quantum mechanics is not enough. An additional physical criterion is required to solve the pathological situation explained above (for instance in the case of the partition function $Z$ of the polymer harmonic oscillator) and such that allows the elimination of most of the eigenstates of the Hamiltonian, just leaving a countable number of them. Here we present a different scenario in which the non-separability of the Hilbert space of the polymer harmonic oscillator plays a non-trivial role. 
 
 Consider for instance the Polymer (Fourier) quantization of the real scalar field given in \cite{Viqar}. In this model, the quantum harmonic oscillator of each Fourier mode $\vec{k}$ of the free scalar field in a flat Minkowski spacetime is replaced by its polymer analog, i.e., by a polymer harmonic oscillator. Formally, the Hilbert space of this quantum field theory can be written as $\prod_{\vec{k}} {\cal H}_{(\vec{k}), poly}$, where ${\cal H}_{(\vec{k}), poly}$ is the polymer Hilbert space with frequency $\omega_{\vec{k}} := \sqrt{\vec{k}^2 + m^2}$ and $m$ is the mass of the free scalar field. A one particle state of the field is given by $| \vec{k}, n_{\vec{k}}, \lambda_{\vec{k}} \rangle $ where $n_{\vec{k}}$ labels the allowed band energy and $\lambda_{\vec{k}}$ parametrized the state in the band $n_{\vec{k}}$. If we consider the polymer vacuum state \cite{Viqar} given by $\prod_{\vec{k}} | \vec{k}, n_{\vec{k}} = 0, \lambda_{\vec{k}} = 0 \rangle$, then transitions of the form $\prod_{\vec{k}} | \vec{k}, n_{\vec{k}} = 0, \lambda_{\vec{k}} = \lambda \delta_{\vec{k}, \tilde{\vec{k}}} \rangle \rightarrow \prod_{\vec{k}} | \vec{k}, n_{\vec{k}} = 0, \lambda_{\vec{k}} = 0 \rangle$ will emit a quantum polymer particle with energy given by 
 \begin{equation}
 E^{\lambda}_0 = \frac{\omega_{\vec{\tilde{k}}}}{2} \left[ 1 - 4 \sqrt{ \frac{M_{\star}}{\pi \, \omega_{ \tilde{\vec{k}}} } } \cos\left( 2 \pi \lambda M^{1/2}_\star \right) e^{-\frac{8 M_{\star}}{\omega_{\tilde{\vec{k}}}} } \right], \label{TamplitudeField}
 \end{equation}
 \noindent as can be seen from the expression (\ref{InstEigenvalue}). Here $M_\star$ stands for the fundamental scale associated with the polymer quantization of the real scalar field $\phi$ (see \cite{Viqar} for details) and is analog to the lattice parameter $\mu$ for the mechanical system. Notice in this example that $\lambda$ turns out to be restricted globally $\lambda_{\vec{k}} \in [0, M^{-\frac{1}{2}}_\star )$, thus, the expression (\ref{TamplitudeField}) can be used to fix bounds in the parameter $M_\star$.

\section*{ACKNOWLEDGMENTS}
We would like to thank Fernando Barbero for useful comments. 
 Angel Garcia-Chung acknowledges the total support from DGAPA-UNAM fellowship. The authors acknowledge partial support from CONACYT project {\bf 237503} and DGAPA-UNAM grant IN {\bf 103716}.


\section{Appendix: Path integral summation} \label{PATHSUM}

Consider the zeroth Hamiltonian amplitude for a fixed value of the parameter $\lambda$ given by 
\begin{equation}
\langle p_f , t_f | p_i, t_i \rangle^{(\lambda)} =\langle p_f | p_i \rangle^{(\lambda)}  =  \left[ \prod^{N}_{n=1} \int^{+\pi p_c}_{- \pi p_c} \frac{dp_n}{2 \pi p_c} \right] \prod^{N+1}_{j=1} \langle p_j  | p_{j-1} \rangle^{(\lambda)} . \label{ZAmplitude}
\end{equation} 

Each infinitesimal amplitude can be written as
\begin{equation}
\langle p_j  | p_{j-1} \rangle^{(\lambda)}  = \frac{1}{\mu} \sum_{n_j} e^{2 \pi i n_j \frac{\lambda}{\mu}} \int^{+\infty}_{- \infty} d x_j \, e^{i x_j (p_j - p_{j-1} - 2 \pi n_j p_c)/ \hbar}.
\end{equation}

Combining these results, the amplitude (\ref{ZAmplitude}) takes the form
\begin{equation}
\langle p_f , t_f | p_i, t_i \rangle^{(\lambda)}  =   \prod^{N}_{n=1} \left[ \int^{+\pi p_c}_{- \pi p_c} \frac{dp_n}{2 \pi p_c} \right] \prod^{N+1}_{j=1}\left[ \sum_{n_j} e^{2 \pi i \, n_j \frac{\lambda}{\mu}} \int^{+\infty}_{- \infty}  \frac{d x_j}{\mu} \, e^{i x_j (p_j - p_{j-1} - 2 \pi n_j p_c)/ \hbar} \right]. 
\end{equation} 

Let us now consider the first integral, i.e., the integral in the variable $p_1$ in the previous expression. It is formed with two sums and two integrations in $x_1$ and $x_2$
\begin{equation}
\int^{+\pi p_c}_{- \pi p_c} \frac{dp_1 }{2 \pi p_c} \left[ \sum_{n_1} e^{2 \pi i \, n_1 \frac{\lambda}{\mu}} \int^{+\infty}_{- \infty}  \frac{d x_1}{\mu} \, e^{i x_1 (p_1 - p_{i} - 2 \pi n_1 p_c)/ \hbar} \right] \left[ \sum_{n_2} e^{2 \pi i \, n_2 \frac{\lambda}{\mu}} \int^{+\infty}_{- \infty}  \frac{d x_2}{\mu} \, e^{i x_2 (p_2 - p_1 - 2 \pi n_2 p_c)/ \hbar} \right]. 
\end{equation} 

We can move the integration in $p$ inside this expression together with one of the summation as
\begin{equation}
\frac{1}{\mu^2} \sum_{n_2} e^{2 \pi i \, n_2 \frac{\lambda}{\mu}} \int^{+\infty}_{- \infty}  d x_1 \, \int^{+\infty}_{- \infty}  d x_2  e^{  i [\frac{ x_2 (p_2 - 2 \pi n_2 p_c) - x_1 p_i}{\hbar} ] } \int^{+\pi p_c}_{- \pi p_c} \frac{dp_1 }{2 \pi p_c} e^{i \frac{ (x_1 - x_2) p_1 }{\hbar} }  \sum_{n_1} e^{2 \pi i \, n_1 \frac{\lambda}{\mu}}  e^{-i \frac{2 \pi n_1 x_1 p_c }{\hbar} }. \label{TWOS}
\end{equation}

Changing variables $\tilde{p}_1 = p_1 - n_1 2 \pi p_c$ yields for the last integral
\begin{equation}
 \int^{+\pi p_c}_{- \pi p_c} \frac{dp_1 }{2 \pi p_c} e^{i \frac{ (x_1 - x_2) p_1 }{\hbar} }  \sum_{n_1} e^{2 \pi i \, n_1 \frac{\lambda}{\mu}}  e^{-i \frac{2 \pi n_1 x_1 p_c }{\hbar} } = \sum_{n_1} e^{- 2 \pi i n_1 \frac{\lambda}{\mu}} e^{2 \pi i x_2 n_1 p_c /\hbar}  \int^{(2 n_1 + 1) \pi p_c}_{(2 n_1 -1) \pi p_c} \frac{d \tilde{p}_1 }{2 \pi p_c} e^{i \frac{(x_1 - x_2 ) \tilde{p}_1}{\hbar}}.
\end{equation}

Substituting this result in (\ref{TWOS}) we obtain
\begin{equation}
\frac{1}{\mu^2} \sum_{n_2} e^{2 \pi i \, n_2 \frac{\lambda}{\mu}} \int^{+\infty}_{- \infty}  d x_1 \, \int^{+\infty}_{- \infty}  d x_2  e^{  i [\frac{ x_2 (p_2 - 2 \pi n_2 p_c) - x_1 p_i}{\hbar} ] } \sum_{n_1} e^{- 2 \pi i n_1 \frac{\lambda}{\mu}} e^{2 \pi i x_2 n_1 p_c /\hbar}  \int^{(2 n_1 + 1) \pi p_c}_{(2 n_1 -1) \pi p_c} \frac{d \tilde{p}_1 }{2 \pi p_c} e^{i \frac{(x_1 - x_2 ) \tilde{p}_1}{\hbar}}.
\end{equation}

Let us now rewrite this last expression in the form
\begin{equation}
 \int^{+\infty}_{- \infty}\frac{d x_1}{\mu} \int^{+\infty}_{- \infty} \frac{d x_2}{\mu} e^{  i \frac{ (x_2 p_2  - x_1 p_i)}{\hbar} }  \sum_{n_2, n_1} e^{2 \pi i \, (n_2 - n_1) \frac{\lambda}{\mu}} e^{ - i \frac{  2 \pi (n_2 - n_1) p_c x_2 }{\hbar} }  \int^{(2 n_1 + 1) \pi p_c}_{(2 n_1 -1) \pi p_c} \frac{d \tilde{p}_1 }{2 \pi p_c} e^{i \frac{(x_1 - x_2 ) \tilde{p}_1}{\hbar}},
\end{equation}
\noindent and redefine the summation label $\tilde{n}_2 = n_2 - n_1$. This gives
\begin{equation}
 \int^{+\infty}_{- \infty}\frac{d x_1}{\mu} \int^{+\infty}_{- \infty} \frac{d x_2}{\mu} e^{  i \frac{ (x_2 p_2  - x_1 p_i)}{\hbar} }  \sum_{\tilde{n}_2, n_1} e^{2 \pi i \, \tilde{n}_2 \frac{\lambda}{\mu}} e^{ - i \frac{  2 \pi \tilde{n}_2 p_c x_2 }{\hbar} }  \int^{(2 n_1 + 1) \pi p_c}_{(2 n_1 -1) \pi p_c} \frac{d \tilde{p}_1 }{2 \pi p_c} e^{i \frac{(x_1 - x_2 ) \tilde{p}_1}{\hbar}}.
\end{equation}

The summations can now be separated again as
\begin{equation}
 \int^{+\infty}_{- \infty}\frac{d x_1}{\mu} \int^{+\infty}_{- \infty} \frac{d x_2}{\mu} e^{  i \frac{ (x_2 p_2  - x_1 p_i)}{\hbar} }  \sum_{\tilde{n}_2} e^{2 \pi i \, \tilde{n}_2 \frac{\lambda}{\mu}} e^{ - i \frac{  2 \pi \tilde{n}_2 p_c x_2 }{\hbar} }  \sum_{ n_1} \int^{(2 n_1 + 1) \pi p_c}_{(2 n_1 -1) \pi p_c} \frac{d \tilde{p}_1 }{2 \pi p_c} e^{i \frac{(x_1 - x_2 ) \tilde{p}_1}{\hbar}}. \label{LTWOS}
\end{equation}

The last summation, together with the integral in $\tilde{p}_1$ can be written as
\begin{equation}
 \sum_{ n_1} \int^{(2 n_1 + 1) \pi p_c}_{(2 n_1 -1) \pi p_c} \frac{d \tilde{p}_1 }{2 \pi p_c} e^{i \frac{(x_1 - x_2 ) \tilde{p}_1}{\hbar}} = \int^{+ \infty}_{- \infty} \frac{d \tilde{p}_1 }{2 \pi p_c} e^{i \frac{(x_1 - x_2 ) \tilde{p}_1}{\hbar}}.
\end{equation}

Inserting this result in (\ref{LTWOS}) we obtain
\begin{equation}
 \int^{+\infty}_{- \infty}\frac{d x_1}{\mu} \int^{+\infty}_{- \infty} \frac{d x_2}{\mu} e^{  i \frac{ (x_2 p_2  - x_1 p_i)}{\hbar} }  \sum_{\tilde{n}_2} e^{2 \pi i \, \tilde{n}_2 \frac{\lambda}{\mu}} e^{ - i \frac{  2 \pi \tilde{n}_2 p_c x_2 }{\hbar} }  \int^{+ \infty}_{- \infty} \frac{d \tilde{p}_1 }{2 \pi p_c} e^{i \frac{(x_1 - x_2 ) \tilde{p}_1}{\hbar}},
\end{equation}
\noindent and redefining $\tilde{n}_2 = n_2$ and $\tilde{p}_1 = p_1$ gives the more familiar form
\begin{equation}
 \sum_{n_2} e^{2 \pi i \, n_2 \frac{\lambda}{\mu}}  \int^{+ \infty}_{- \infty} \frac{d p_1 }{2 \pi p_c} \int^{+\infty}_{- \infty}\frac{d x_1}{\mu} e^{ i \frac{ x_1 (p_1 - p_i) }{\hbar}} \int^{+\infty}_{- \infty} \frac{d x_2}{\mu}  e^{  i \frac{x_2 (p_2 - p_1 - 2 \pi n_2 p_c)}{\hbar}  } .
\end{equation}

Let us summarize. For each integral in $p_j$ there are two summations $n_j$ and $n_{j+1}$. The one with label $n_j$ is absorbed in the expansion of the interval for the momentum $p_j$ together with the phase in $\frac{\lambda}{\mu}$. Due to we are dealing with $N$ integrals in $p$ and $N+1$ summations, the summation with label $n_{N+1}$ will still remain. Therefore, the amplitude for the zeroth Hamiltonian takes the form
\begin{equation}
\langle p_f , t_f | p_i, t_i \rangle^{(\lambda)}  =  \sum_{n} e^{2 \pi i \, n \frac{\lambda}{\mu}} \prod^{N}_{n=1} \left[ \int^{+\infty}_{- \infty} \frac{dp_n}{2 \pi p_c} \right] \prod^{N+1}_{j=1}\left[  \int^{+\infty}_{- \infty}  \frac{d x_j}{\mu} \, e^{i x_j (p_j - p_{j-1} - 2 \pi n \delta_{j, N+1} p_c)/ \hbar} \right]. 
\end{equation} 

The changes of variable involved in this result does not affect the form of the Hamiltonian due to $H(x_j, p_j) = H(x_j, p_j + 2 \pi n_j p_c)$, hence, it can be easily extended to the polymer Hamiltonian yielding
\begin{equation}
\langle p_f , t_f | p_i, t_i \rangle^{(\lambda)}  =  \sum_{n} e^{2 \pi i \, n \frac{\lambda}{\mu}} \prod^{N}_{n=1} \left[ \int^{+\infty}_{- \infty} \frac{dp_n}{2 \pi p_c} \right] \prod^{N+1}_{j=1}\left[  \int^{+\infty}_{- \infty}  \frac{d x_j}{\mu} \, e^{i x_j (p_j - p_{j-1} - 2 \pi n \delta_{j, N+1} p_c)/ \hbar - i \epsilon H^{(\mu)}_{poly}(x_j, p_j)/\hbar} \right]. 
\end{equation} 

\section{Appendix: Solution of the Quantum fluctuation equation} \label{SQFEq}

In this appendix we are going to summarize the solution to the eigenvalue problem of the equation (\ref{PolyInst}). We proceed along the notes given in \cite{Kazama}. To begin with, consider the following potential 
\begin{equation}
U(\tau) = 1 - A \, \mbox{Sech}^2(B \, \omega \Delta \tau),
\end{equation}
\noindent where $A$ and $B$ are arbitrary real constants. The equation for the quantum fluctuation in this potential takes the form
\begin{equation}
- \frac{d^2}{d\tau^2}f_n(\tau) + \omega^2 U(\tau) f_n(\tau) = \epsilon_n \, m \, \omega^2 \, f_n(\tau). \label{FEQGen}
\end{equation}
 Let us define the dimensionless variable  
\begin{equation}
\xi := \mbox{tanh} (B\, \omega \Delta \tau),
\end{equation}
\noindent and let us write the equation (\ref{FEQGen}) in terms of $\xi$
\begin{equation}
\frac{\partial}{\partial \xi} \left[ \left( 1 - \xi^2 \right) \frac{\partial f_n(\xi)}{\partial \xi}\right] + \left\{ \frac{A}{B^2} - \frac{(1 - m \epsilon_n)}{B^2 \,(1 - \xi^2)} \right\} f_n(\xi) = 0. \label{ALPoly}
\end{equation}
This is the equation of the associated Legendre polynomials with $s=1$ and $e^2 := 1 - m \epsilon$, see \cite{Landau} for details. This equation can be written in the form of an hypergeometric equation if we define the new function $f_n := (1 - \xi^2)^C \chi_n(\xi)$, where $C$ is an arbitrary constant. We additionally consider another change of variable and define $\xi = 1 - 2z$. As a result, the equation (\ref{ALPoly}) turns into
\begin{equation}
z(1 - z) \chi''_n + [1 + 2 C - (2 + 4 C) z] \chi'_n + \left\{ \frac{A}{B^2} - 2 C - 4 C^2 + \frac{4C^2 + \frac{m\epsilon_n - 1}{B^2}}{4z (1 - z)} \right\} \chi_n = 0. \label{HEq}
\end{equation}

The relation between these changes of variable is given as 
\begin{eqnarray}
&&\tau \rightarrow + \infty \quad  \Rightarrow \quad \xi \rightarrow 1 \quad \Rightarrow \quad z \rightarrow 0 , \quad \mbox{and} \quad \tau \rightarrow - \infty  \quad  \Rightarrow \quad \xi \rightarrow -1 \quad \Rightarrow \quad z \rightarrow 1 . \nonumber
 \end{eqnarray}
 Let us impose in the equation (\ref{HEq}) the following condition for $C$
\begin{equation}
4C^2 + \frac{m\epsilon_n - 1}{B^2} =0,
\end{equation}
\noindent which removes the $z-$dependence of the last coefficient in (\ref{HEq}). With this condition, the hypergeometric equation takes the following form 
\begin{eqnarray}
z(1 - z) \frac{d^2 \chi_n}{dz^2} + [\gamma - (\alpha + \beta + 1) z] \frac{d\chi_n}{dz} - \alpha \beta \chi_n = 0, \label{HGF}
\end{eqnarray}
\noindent where the parameters $\alpha$, $\beta$ and $\gamma$ are defined as 
\begin{equation}
\alpha = \frac{1 + 4C}{2} - \sqrt{\frac{1}{4} + \frac{A}{B^2}}, \qquad \beta = 1+4C - \alpha, \qquad \gamma = 1 + 2C.  \label{PDefinition}
\end{equation}

The first solution of (\ref{HGF}) is the hypergeometric function $F(\alpha, \beta, \gamma; z)$, which is a series in the variable $z$ given by
\begin{equation}
F(\alpha, \beta, \gamma; z) = 1 + \frac{\alpha \, \beta}{\gamma} \frac{z}{1 !} + \frac{\alpha (\alpha +1) \, \beta (\beta + 1)}{\gamma (\gamma + 1)} \frac{z^2}{2 !} + \cdots \label{HGeomF}
\end{equation}
\noindent and is regular at $z=0$ when $\gamma \neq 0, -1, -2, \dots$. The other independent solution of (\ref{HGF}) is given by
\begin{equation}
z^{1 - \gamma} F( \alpha - \gamma + 1, \beta - \gamma +1, 2 - \gamma; z) ,
\end{equation}
\noindent and is singular at $z=0$. 

Consider now a discrete spectrum within in the interval $- \frac{1}{m} < \epsilon_n < +\frac{1}{m}$. In this case, the constant $C$ acquires real values $ C = \pm \frac{1}{2 B} \sqrt{1 - m \epsilon_n}$. The sign is selected by imposing the condition $f_n(\Delta \tau \rightarrow + \infty) = 0$ which implies that $C$ is actually $C =  \frac{1}{2 B} \sqrt{1 - m \epsilon_n}$. The other limit $f_n(\Delta \tau \rightarrow - \infty) = 0$ implies that the series in $F$ must be truncated due to the singularity of the hypergeometric function when $z =1$. This is only possible if $\alpha$ or $\beta$ ((\ref{HGeomF}) is symmetric in both parameters) takes a negative natural number value, in other words, $\alpha = - n$. Let us consider the potential (\ref{FQPot}) with the values $A=2$ and $B=1$. The expression for $\alpha$ given in (\ref{PDefinition}) yields the condition
\begin{equation}
\sqrt{1-m \epsilon_n} -1 = -n \quad \Rightarrow  \quad n=0, \qquad \epsilon^d_{n=0} =0. 
\end{equation}
There is only one value $\epsilon^d_0 = 0$ for the discrete spectrum and it is also a zeroth mode \cite{Coleman}. The eigenfunction of this mode is 
\begin{equation}
f_0(\tau) := \sqrt{\frac{\hbar m \omega}{2}} \cosh^{-1}(\omega \Delta \tau),
\end{equation}
\noindent and it is normalized on the infinite time interval. Combining the zeroth mode $f_0(\tau)$ with the derivative of the instanton solution given in (\ref{SolOP}) and the value of the action on this solution (\ref{IVAction}) the zeroth mode can be written as
\begin{equation}
f_0(\tau) = \sqrt{\frac{\hbar}{\omega^2 S_0}} \left( \frac{dP^+}{d\tau} \right). \label{INSTZeroth}
\end{equation}
This expression will be used in the calculation of the ratio between the penetration amplitude during the renormalization procedure.

 It is worth to mention that in the case of the double well system, the parameters $A$, $B$ takes the values $A=3/2$ and $B=1/2$. Inserting these values in the condition for the discrete spectrum gives
\begin{equation}
-2 + 2 \sqrt{1-m\epsilon_n} = -n,
\end{equation} 
\noindent which yields the additional discrete eigenvalue $\epsilon_1 = \frac{3}{4m}$. For more general values of $A$ and $B$ the condition reads as
\begin{equation}
1-A \leq \epsilon_m < 1, \quad \mbox{with} \quad \epsilon_n = \frac{1}{m} \left\{ 1 - B^2 \left[ \sqrt{\frac{1}{4} + \frac{A}{B^2}} - \left( n + \frac{1}{2}\right)\right]^2\right\}.
\end{equation}

Let us now consider the continuous spectrum. As we already mentioned, in the continuous spectrum the eigenvalues are arbitrary real numbers such that $\epsilon > 1/m$. Therefore, each eigenvalue can be labelled by a real `momentum'  $\tilde{p}$ such that 
\begin{equation}
\epsilon_{\tilde{p}} = \frac{1}{m} + \frac{\tilde{p}^2}{m\omega^2} . \label{QFMomentum}
\end{equation}
The absent of a barrier for this values of $\epsilon_{\tilde{p}}$ allow us to discard possible reflections of the quantum fluctuation as it travels from $\tau = -\infty$ to $\tau = +\infty$. Moreover, this values of $\epsilon_{\tilde{p}}$ implies that $C$ becomes imaginary $C = \pm \frac{i}{2} \frac{ \tilde{p}}{\omega}$. In order to study solutions with asymptotic behavior given by $e^{i \tilde{p} \tau}$, we consider only the constant $C = \frac{i\,\tilde{p}}{2\, \omega} = \frac{i}{2} \sqrt{|1-m\epsilon_{\tilde{p}} |}$. 

 Recall now that $\Delta \tau \rightarrow + \infty$ implies $z \rightarrow 0$. By taking this consideration, the unique stable solution will be of the form 
\begin{equation}
f(\tau) = \left[ \frac{e^{+\omega \Delta \tau} + e^{-\omega \Delta \tau} }{2} \right]^{i \sqrt{|1-m\epsilon_{\tilde{p}}|}} F(-1 + 2C, 2 + 2C, 1 + 2C; z) , \label{SolutionCS}
\end{equation}
\noindent and notice that for long times it takes the form
\begin{equation}
 f(\tau \rightarrow + \infty) \approx \left( \frac{e^{\omega \Delta \tau} }{2} \right)^{i \sqrt{|1 - m \epsilon_{\tilde{p}}|} }.
 \end{equation}

We now use the following property of the hypergeometric function \cite{Landau}
\begin{eqnarray}
F(\alpha, \beta, \gamma; z) &=& \frac{\Gamma(\gamma) \Gamma(\gamma - \alpha - \beta)}{\Gamma(\gamma - \alpha) \Gamma(\gamma - \beta)}  F(\alpha, \beta,  \alpha + \beta +1 - \gamma; 1 - z) + \nonumber \\
&& + \frac{\Gamma(\gamma) \Gamma( \alpha + \beta - \gamma)}{\Gamma( \alpha) \Gamma( \beta)} (1 - z)^{\gamma - \alpha - \beta}  F(\gamma  - \alpha, \gamma - \beta,  \gamma - \alpha - \beta + 1 ; 1 - z) ,
\end{eqnarray}
\noindent and rewrite the function (\ref{SolutionCS}) as
\begin{eqnarray}
f(\tau) &=&  (1 - \xi^2)^C \frac{\Gamma(1 + 2C) \Gamma(-2 C)}{\Gamma(2) \Gamma(-1)}  F(-1+2C, 2+2C,1+2C; 1 - z) + \nonumber \\
&& + (1-\xi^2)^C (1-z)^{-2C} \frac{\Gamma(1+2C) \Gamma(2C)}{\Gamma(-1+2C) \Gamma(2+2C)} F(2,-1,1-2C ; 1 - z) .
\end{eqnarray}

The term $\Gamma(-1) \rightarrow + \infty$ and therefore the first contribution is not considered. Notice that the limit $\tau \rightarrow - \infty$ gives $z \rightarrow 1$ and it implies that $F(-1+2C, 2+2C,1+2C; 0)$, which is a regular function. When $\tau \rightarrow - \infty$ the other contribution can be written as
\begin{eqnarray}
f(\tau \rightarrow - \infty) \approx \left( \frac{e^{\omega \Delta \tau} }{2} \right)^{i \sqrt{|1 - m \epsilon_{\tilde{p}}|} } \frac{\Gamma(1+2C) \Gamma(2C)}{\Gamma(-1+2C) \Gamma(2+2C)} = \left( \frac{e^{\omega \Delta \tau} }{2} \right)^{i \sqrt{|1 - m \epsilon_{\tilde{p}}|} } e^{i \delta_{\tilde{p}}},
\end{eqnarray}
\noindent where the phase $\delta_{\tilde{p}}$ is given by
\begin{equation}
\delta_{\tilde{p}} = \pi + 2 \tan^{-1} \left( \sqrt{| 1- m \epsilon_{\tilde{p}} |} \right)  .
\end{equation}
This is the phase we were looking for and then the next (and last) step in this analysis is to impose the `temporal box' boundary conditions. These conditions turn the continuum spectrum into a discrete one by demanding that the solutions are null in a finite time interval $(\tau_i, \tau_f)$, i.e., $f_{\epsilon}(\tau_i) = f_{\epsilon}(\tau_f) =0$. Using the asymptotic form of the solutions, we obtain the following condition
\begin{equation}
\frac{\left( \frac{e^{ \frac{\omega\tau_0}{2}} }{2} \right)^{i \sqrt{|1 - m \epsilon_{\tilde{p}}|} }}{ \left( \frac{e^{ \frac{\omega \tau_0}{2}} }{2} \right)^{i \sqrt{|1 - m \epsilon_{\tilde{p}}|} } e^{i \delta_{\tilde{p}}} } = \pm 1.
\end{equation}
As a result, due to the relation (\ref{QFMomentum}) the momentum $\tilde{p}$ will also be discrete. In this case, it takes the form\begin{equation}
\tilde{p}_n = \frac{n \, \pi + \delta_{\tilde{p}}}{\tau_0} = \frac{(n+1)\pi}{\tau_0} + \frac{2}{\tau_0} \tan^{-1}\left(\frac{\tilde{p}_n}{\omega}\right).
\end{equation} 
This is a transcendental equation for $\tilde{p}_n$. Each value of $\tilde{p}_n$ gives an eigenvalue $\epsilon_n$ using (\ref{QFMomentum}). This expression will used in the regularization of the polymer harmonic oscillator amplitude. 



\begin{thebibliography}{9}
\bibitem{ShadowsStates} A.~Ashtekar, S.~Fairhurst and J.~L.~Willis, ``Quantum gravity, shadow states, and quantum mechanics,'' Class.\ Quant.\ Grav.\  {\bf 20}, 1031 (2003). http://cgpg.gravity.psu.edu/archives/ thesis/2004/willis thesis.pdf
\bibitem{PQMCLimit} A.~Corichi, T.~Vukasinac and J.~A.~Zapata, ``Polymer Quantum Mechanics and its Continuum Limit,'' Phys.\ Rev.\ D {\bf 76}, 044016 (2007). 
\bibitem{HPHSPQM} A.~Corichi, T.~Vukasinac and J.~A.~Zapata, ``Hamiltonian and physical Hilbert space in polymer quantum mechanics,'' Class.\ Quant.\ Grav.\  {\bf 24}, 1495 (2007).
\bibitem{LQProgram1} A.~Ashtekar and J.~Lewandowski ``Background independent quantum gravity: A status report,'' Class. Quant. Grav. {\bf 21}, R53 (2004); 
\bibitem{LQProgram2} C.~Rovelli, ``Quantum Gravity,'' (Cambridge University Press, 2004).
\bibitem{LQProgram3} T.~Thiemann, ``Modern Canonical Quantum General Relativity,'' (Cambridge University Press, 2007).
\bibitem{Bratelli} R.~V.~Kadison and J.~R.~Ringrose, ``Fundamentals of the Theory of Operator Algebras: Elementary Theory,'' Vol I, (Academic Press, 1983).


\bibitem{vonNeumann} J.~von Neumann, `Die Eindeutigkeit der Schr\"odingerschen operatoren,'' Math.\  Ann. {\bf 104}, (1931); Collected Works, Vol. 2, No. 7.
\bibitem{Strocchi1} F.~Acerbi, G.~Morchio and F.~Strocchi, ``Infrared singular fields and nonregular representations of canonical commutation relation algebras,'' J.\ Math.\ Phys. {\bf 34}, 899 (1993).
\bibitem{Strocchi2} F.~Cavallaro, G.~Morchio and F.~Strocchi, ``A generalization of the Stone-von Neumann theorem to non-regular representations of the CCR-algebra,'' Lett.\ Math.\ Phys. {\bf 47}, 307 (1999).
\bibitem{Halvorson} H.~Halvorson, ``Complementarity of Representations in quantum mechanics,'' Studies in History and Philosophy of Modern Physics {\bf 35}, 45 (2004).
\bibitem{Anakin} E.~Flores-Gonz\'alez, H.~A.~Morales-T\'ecotl and J.~D.~Reyes, ``Propagators in Polymer Quantum Mechanics,'' Annals Phys  {\bf 336}, 394 (2013).
\bibitem{ViqarLouko} V. Husain, J. Louko and O. Winkler, ``Quantum gravity and the Coulomb potential,'' Phys.\ Rev.\ D {\bf 76}, 084002 (2007). 
\bibitem{Louko1} G.~Kunstatter, J.~Louko and J.~Ziprick, ``Polymer quantization, singularity resolution and the 1/r2 potential,'' Phys.\ Rev.\ A {\bf 79}, 032104 (2009). 
\bibitem{Louko2} G.~Kunstatter, J.~Louko and A.~Peltola, ``Polymer quantization of the Einstein-Rosen wormhole throat,'' Phys.\ Rev.\ D {\bf 81}, 024034 (2010).
\bibitem{Barbero} J.~F.~Barbero G., J.~Prieto and E.~J.~S.~Villase\~nor, ``Band structure in the polymer quantization of the harmonic oscillator,'' Class.\ Quant.\ Grav.\  {\bf 30}, 165011 (2013)
\bibitem{Abhay} A.~Ashtekar, P.~Singh, ``Loop quantum cosmology: a status report,'' Class.\ Quant.\ Grav. {\bf 28}, 213001 (2011).
\bibitem{Martin} M.~Bojowald, ``Loop quantum cosmology,'' Living Reviews in Relativity {\bf 11}, 66 (2008).

\bibitem{Viqar} G.~M.~Hossain, V.~Husain, and S.~S.~Seahra, ``Propagator in polymer quantum field theory,'' Phys.\ Rev.\ D {\bf 82}, 124032 (2010).
\bibitem{Memo1} G.~Chac\'on-Acosta, E.~Manrique, L.~Dagdug, H.~A.~Morales-T\'ecotl, ``Statistical thermodynamics of polymer quantum systems,'' SIGMA {\bf 7},  110 (2011).
\bibitem{Memo2} E.~Castellanos and G.~Chac\'on-Acosta, ``Polymer Bose Einstein condensates,'' Phys.\ Lett.\ B {\bf 722}, 119 (2013)
\bibitem{Memo3} G.~Chac\'on-Acosta and H.~H.~Hern\'andez-Hern\'andez, ``Polymer quantum effects on compact stars models,'' Int.\ J.\ Mod.\ Phys.\ D {\bf 24}, 1550033 (2015).
\bibitem{HugoyYo}  A.~A.~Garc\'ia-Chung and H.~A.~Morales-T\'ecotl, ``Polymer Dirac field propagator: A model,'' Phys.\ Rev.\ D {\bf 89},  065014 (2014). 
\bibitem{Bohr} C.~Corduneanu, ``Almost Periodic Functions'', (Chelsea Publishing Co. New York, 1989). E.~Hewitt and K.~A.~Ross ``Abstract Harmonic Analysis'',  Vol 1 (Academic Press, New York, 1963).
\bibitem{Velhinho} J.~M.~Velhinho, ``The Quantum configuration space of loop quantum cosmology,''  Class.\ Quant.\ Grav.\  {\bf 24}, 3745 (2007).
 \bibitem{ReedSimon} M.~Reed and B.~Simon, ``Methods of Modern Mathematical Physics,'' Vol 4 (Academic Press, New York, 1978).

\bibitem{Abramovitz} M.~Abramowitz and I.~A.~Stegun, ``Handbook of Mathematical Functions,'' (Dover Publications, New York, 1972).
\bibitem{Campiglia} A.~Ashtekar, M.~Campiglia and A.~Henderson, ``Path Integrals and the WKB approximation in Loop Quantum Cosmology,'' Phys.\ Rev.\ D {\bf 82}, 124043 (2010).
\bibitem{ParraDavid} L.~Parra and J.~D.~Vergara, ``Polymer quantum mechanics some examples using path integrals,'' AIP Conference Proceedings, {\bf 1577}, 269 (2014).
\bibitem{Hugo} H.~A.~Morales-T\'ecotl, D.~H.~Orozco-Borunda and S.~Rastgoo, ``Polymer quantization and the saddle point approximation of partition functions,'' Phys.\ Rev.\ D {\bf 92}, 104029 (2015).

\bibitem{Cycon} H.~L.~Cycon, R.~G.~Froese, W.~Kirsch and B.~Simon, ``Schr\"odinger operators with application to quantum mechanics and Global Geometry,'' (Springer-Verlag Berlin Heidelberg, 1987).
\bibitem{Araki} ``Topics in the theory of Schr\"odinger operators,'' World Scientific Publishing Co. 2004

\bibitem{Tomasz1} A.~Kreienbuehl and T.~Pawlowski, ``Singularity resolution from polymer quantum matter,'' Phys.\ Rev.\ D {\bf 88}, 043504 (2013).
\bibitem{Tomasz2} W.~Kaminski, J.~Lewandowski and T.~Pawlowski, ``Quantum constraints, Dirac observables and evolution: Group averaging versus Schrodinger picture in LQC,'' Class.\ Quant.\ Grav.\  {\bf 26}, 245016 (2009).
\bibitem{Barbero2} J.~F.~Barbero G., T.~Pawlowski and E.~J.~S.~Villasenor, ``Separable Hilbert space for loop quantization,'' Phys.\ Rev.\ D {\bf 90}, 067505 (2014).

\bibitem{TheoremSH} M.~A.~Shubin, ``Theorems on the coincidence of the spectra of pseudodifferential almost-periodic operators in the spaces $L^2(\mathbb{R}^n)$ and $B^2(\mathbb{R}^n)$.'' Sib. Math. J. {\bf 17}, 158, (1976). M.~Burnat, ``The spectral properties of the Schr\"odinger operator in nonseparable Hilbert spaces,'' Banach Center Publications (1982). J.~Herczy\'nski ``Schr\"odinger operators with almost periodic potentials in nonseparable Hilbert spaces.'' Banach Center Publications 121. (1987). 

\bibitem{BSimon} J.~Avron and B.~Simon, ``The asymptotics of the gap in the Mathieu equation,'' Annals of Physics {\bf 134}, 76 (1981).

\bibitem{Kittel} C.~Kittel, ``Introduction to Solid State,'' (John Wiley $\&$ Sons, 1966).

\bibitem{Coleman} S.~Coleman, ``Aspects of Symmetry,'' (Cambridge University Press, 1985).  

\bibitem{Rajamaran} R.~Rajaraman, ``Solitons and Instantons,'' (Elsevier, Amsterdam, 1982).

\bibitem{ABCInstanton} A.~I.~Vainshtein, V.~I.~Zakharov, V.~A.~Novikov and M.~A.~Shifman, `ABC's of Instantons,''  Sov.\ Phys.\ Usp.\  {\bf 25}, 195 (1982).

\bibitem{Kazama} http://hep1.c.u-tokyo.ac.jp/~kazama/QFT/instanton2.pdf

\bibitem{Paradis} F.~Paradis, H.~Kroger, G.~Melkonyan and K.~J.~M.~Moriarty, ``New vista on quantum tunneling and instantons,'' Phys.\ Rev.\ A {\bf 71}, 022106 (2005)

\bibitem{Chalbaud} E.~Chalbaud, J.~P.~Gallinar and G.~Mata, ``The quantum harmonic oscillator on a lattice,'' J. Phys. A {\bf 19}, L385 (1986).
\bibitem{Wilson} K.~G.~Wilson, ``Confinement of Quarks,'' Phys.~Rev.~D {\bf 10}, 2445 (1974).
\bibitem{Jan} J.~Smit,``Introduction to Quantum Fields on a Lattice: a robust mate,'' (Cambridge University Press, 2002).


\bibitem{Kleinert} H.~Kleinert, ``Path Integrals in Quantum Mechanics, Statistics, Polymer Physics, and Financial Markets,'' (World Scientific, Singapore, 2009)

\bibitem{BSimon2} M.~Stone, ``Periodic Vacua and Functional Integrals: A Toy Model,'' Phys.\ Rev.\ D {\bf 18}, 4752 (1978).
\bibitem{Chojnacki} W.~Chojnacki, ``Eigenvalues of almost periodic Schr\"odinger operators in $L^2(b\mathbb {R})$ are at most double,'' Lett.\ Math.\ Phys. {\bf 22}, (1991).

\bibitem{Landau} L.~D.~Landau and E.~M.~Lifshitz, ``Quantum Mechanics,'' (Pergamon Press , 1965). 

\end{thebibliography}
\end{document}